# Diffractive production of dijets at HERA

M. Diehl

Department of Applied Mathematics and Theoretical Physics,
University of Cambridge, Cambridge CB3 9EW, England

**Abstract**

We investigate the production of a quark-antiquark pair in diffractive photon-proton scattering, approximating soft pomeron exchange by the exchange of two nonperturbative gluons. In deep inelastic scattering at HERA, events with two jets and the scattered proton in the final state are predicted to be observable, with an important contribution from charm production. For photoproduction of light quark jets with high transverse momentum we find that both exchanged gluons must have a large invariant mass, so that the cross section is very small, whereas for charm quarks it is quite appreciable. From our calculation we also extract the quark structure function of the pomeron for the scaling variable $z$ no too close to 0 or 1, finding a strong flavour dependence and a behaviour somewhat harder than $z(1-z)$ for light quarks.

# 1 Introduction

Despite the great progress in describing strong interactions that has been achieved since the advent of QCD, processes that involve large distances, or small momentum transfers, are far from being well understood at present. Since the strong coupling becomes large at small momentum scales, perturbation theory cannot be directly applied in this situation, at least not in a straightforward manner. On the other hand, the phenomenology of soft processes, such as diffractive scattering or dissociation of hadrons, is fairly well described by the exchange of Regge trajectories. Most of these can be associated with mesons or baryons, but not so the trajectory with vacuum quantum numbers, the pomeron. The investigation of its properties and of how it can be described in terms of QCD is a challenge both for experiment and theory.

At HERA, one is effectively colliding protons with real or virtual photons. One can look for collisions where the proton stays intact and loses only a tiny fraction of its momentum (instead of the proton there may also be a low-lying excitation such as $N\pi$ in the final state, cf. [1]), and where the pomeron it has radiated interacts with the photon. For purely kinematic reasons, the centre of mass of the photon-pomeron subreaction is well separated in rapidity from the scattered proton or proton remnant. Since the exchanged object is colourless there is no need for colour rearrangement in the final state which would "fill" this "rapidity gap" with further hadrons. Such a reaction can probe electromagnetic interactions of the pomeron. Given the high energies of the photon-proton collision at HERA, the photon-pomeron system can have an invariant mass large enough to allow for a hard reaction, which is amenable to a perturbative calculation, so that one is at the interface of soft and hard QCD. Events with a large rapidity gap and a jet structure in the final state have indeed been observed at HERA [2] and will be studied in more detail in the next few years.

In this paper we investigate diffractive events with only the scattered proton and a pair of jets in the final state, the jets coming from a quark-antiquark pair produced in the photon-pomeron interaction (fig. 1). We consider both light and heavy flavours for the quarks, and both photoproduction and deep inelastic scattering (DIS). The framework of our calculation is the QCD-inspired model of Landshoff and Nachtmann [3, 4, 5], where the pomeron is approximated by two gluons with a nonperturbative propagator. It successfully reproduces some of the known phenomenological properties of the pomeron and has been used to describe exclusive diffractive $\rho$ production in deep inelastic scattering. The process $p + \gamma \to p + q\bar{q}$ with real photons has previously been treated in this model [6], but there was an error in the calculation. For virtual photons the same reaction has been extensively studied in purely perturbative QCD, i. e. using the perturbative gluon

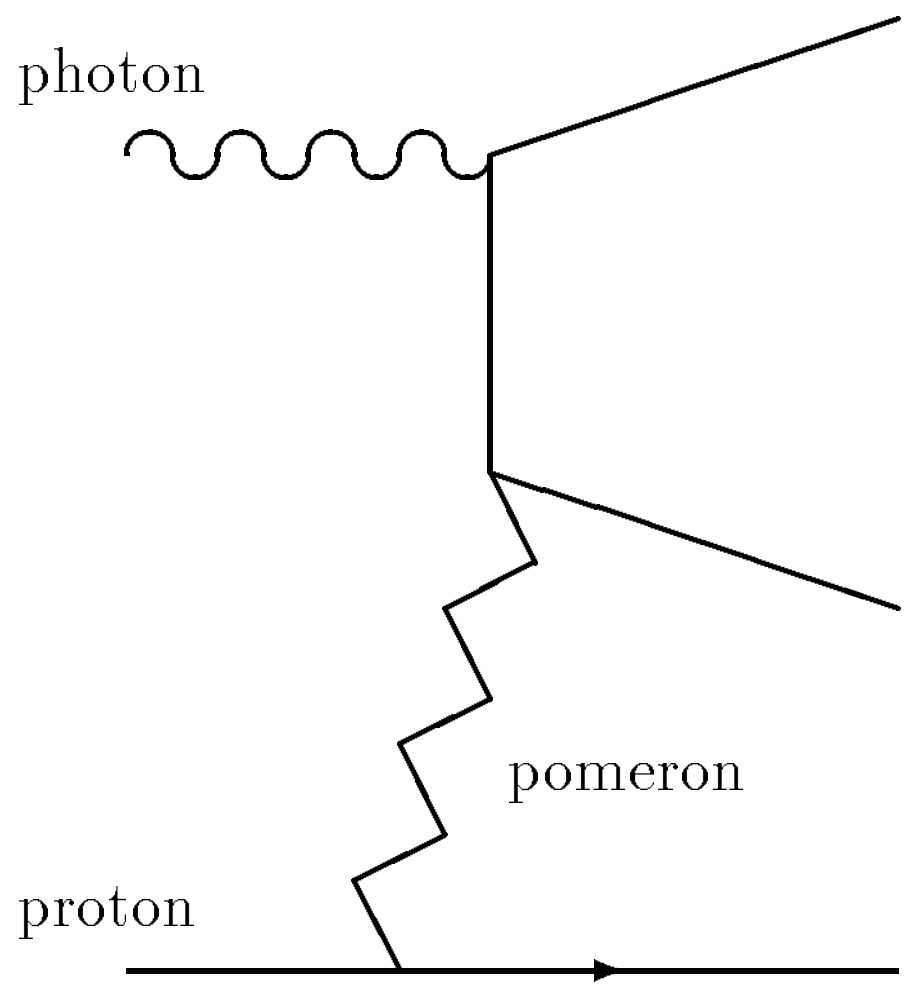

Figure 1: Diffractive production of $q\bar{q}$ by pomeron exchange. The figure corresponds to two diagrams because of the two possible directions of charge flow in the upper quark line.

propagator [7, 8, 9, 10, 11].

In some aspects, the pomeron behaves quite like a particle. Ingelman and Schlein proposed to describe processes in which a pomeron undergoes a hard reaction in terms of a parton structure of the pomeron and introduced the corresponding structure functions [12]. In the parton model for ordinary hadrons the quark structure function is related with the total cross section for photoabsorption, which can be generalised to the quark structure function of the pomeron and the cross section for *diffractive* photoabsorption. We can calculate the latter at Born level if we integrate the cross section for diffractive $q\bar{q}$-production over the transverse quark momentum, and thus estimate the quark structure of the pomeron in our model.

The present paper is organised as follows. In section 2 we describe the kinematics of the process under study and the kinematical limit we are working in. Section 3 briefly recapitulates the Landshoff-Nachtmann (LN) model of the pomeron and section 4 describes the calculation of the cross section for diffractive $q\bar{q}$-production in this model. Our results are presented separately for photoproduction and DIS in the following two sections. In particular, we find for photoproduction of *light* quarks that the exchanged gluons must each have a large invariant mass in order to contribute to the cross section. This is not the case for heavy flavours, and we obtain a cross section for photoproduction of charm which should be observable at HERA (incidentally it is of the same order as the one calculated for light quarks in [6]). In section 7 we calculate and discuss the quark structure function of the pomeron in the LN model. The last section contains our conclusions. In an appendix, we briefly present what the phenomenological model of the pomeron by Donnachie and Landshoff [13, 14] predicts for our reaction, and show that it can be used to calculate the pomeron structure function, but not the production of high-$p_T$ jets.

# 2 Kinematics

We begin with introducing some convenient kinematic variables for the process

$$p(p) + \gamma(q) \to p(p') + q(P) + \bar{q}(P') \ . \tag{1}$$

Let $m$ be the mass of the quark $q$ and $\sqrt{\hat{s}}$ the invariant mass of the quark-antiquark pair. We consider both real and virtual photons, and use the conventional abbreviations

$$Q^2 = -q^2, \ \ W^2 = (p+q)^2, \ \ \nu = p.q \tag{2}$$

as well as $M_p$ for the proton mass.

For the momentum transfer of the proton (i. e. the momentum carried away by the pomeron) we use a Sudakov parametrisation

$$p - p' = \Delta = \xi p + \xi' q + \Delta_T \tag{3}$$

with $\Delta_T$ transverse to both $p$ and $q$. The on-shell conditions $p^2 = p'^2 = M_p^2$ for the proton lead to a quadratic equation in $\xi$ and $\xi'$. Since we are interested in diffractive scattering, the squared momentum transfer $t = \Delta^2$ is small, say $|t| \leq 1\,\mathrm{GeV}^2$, and the ratio $(W^2+Q^2)/(\hat{s}+Q^2)$ is large, i. e. we are working in the Regge limit. As a consequence one finds that $0 < \xi \ll 1$, and that only one of the two solutions for the on-shell constraint is applicable, namely

$$\begin{aligned} \xi' &\approx \frac{\Delta_T^2 - 2\xi M_p^2}{2\nu} \\ t &\approx \Delta_T^2 - \xi^2 M_p^2 \ . \end{aligned} \tag{4}$$

Provided that $\hat{s} \gg |\Delta_T^2|$, one can neglect $\xi'$ compared with $\xi$. One then has

$$\xi = \frac{\hat{s} + Q^2}{W^2 + Q^2} \tag{5}$$

and can interpret $\xi$ as the proton's fractional loss of longitudinal momentum (cf. (3)). Requiring a minimum invariant jet mass $\sqrt{\hat{s}}$ translates into a lower bound for $\xi$. We shall also impose an upper bound of 0.01 on $\xi$ to be reasonably sure that the exchange of a pomeron dominates that of other Regge trajectories [6, 15], even at $t \approx 0$. In most of our numerical applications, the cross section is dominated by smaller $\xi$ and the effect of this bound is rather weak.

In our calculation we shall neglect the transverse momentum $\Delta_T$. To leading order in $\xi^{-1}$ one then has from (3)

$$p \approx p' \approx \xi^{-1}\Delta \ . \tag{6}$$

The proton mass will not appear in our results, since it turns out that terms involving $M_p$ are suppressed at least by $\xi$, such as in (4).

We further introduce the transverse momentum $p_T$ of the quark relative to the photon direction in the $q\bar{q}$ centre of mass. Note that for fixed $\mathbf{p}_T$ and $\hat{s}$ the angle between the quark and the photon can take two different values which sum up to 180°. Interchanging these two values is tantamount to interchanging the quark and antiquark momenta, and due to charge conjugation invariance the squared transition amplitude is the same for both configurations. Summing over them gives a factor of 2 in the phase space element. After integration over the azimuthal angles associated with $p_T$ and $\Delta_T$, one has

$$\frac{d^3\mathbf{p}'}{2p'_0}\frac{d^3\mathbf{P}}{2P_0}\frac{d^3\mathbf{P}'}{2P'_0}\,\delta^4(p'+P+P'-p-q) = \frac{\pi^2}{2}\frac{1}{\hat{s}\sqrt{1-4(\mathbf{p}_T^2+m^2)/\hat{s}}}\,dt\,d\xi\,d\mathbf{p}_T^2 \ . \tag{7}$$

We require the transverse jet momentum to be large, say $|\mathbf{p}_T| \geq 3\,\mathrm{GeV}$ at parton level, so that the jets can be identified experimentally and the produced quarks may be treated perturbatively in our calculation, including the case when $Q^2 = 0$.

# 3 The Landshoff-Nachtmann model

In the LN model [3, 4, 5] the (soft) pomeron is modelled within QCD by the colour singlet part of two-gluon exchange. In contrast to a purely perturbative approach, nonperturbative effects are partly taken into account by allowing for the confinement of the gluons or, more precisely, by working with a nonperturbative gluon propagator. Nonperturbative effects are essential at low squared gluon momentum $l^2$, whereas at hard scales perturbative behaviour dominates. When the perturbative piece, $-g_{\mu\nu}/l^2$ in Feynman gauge, is subtracted at large $l^2$ one remains with a gluon propagator $-g_{\mu\nu}D(l^2)$ which can be related to the gluon condensate of the nonperturbative QCD vacuum. It was found in [3] that in order to reproduce the additive quark rule for total hadronic cross sections $D(0)$ has to be finite. Applied to certain processes, knowledge of the complete form of $D(l^2)$ is not necessary, and only some of its moments are needed. In particular [4, 5, 6][1]

$$\begin{aligned} \int_0^\infty dl^2 [\alpha_s^{(0)} D(-l^2)]^2 &= \frac{9\beta_0^2}{4\pi} \\ \int_0^\infty dl^2\, l^2 [\alpha_s^{(0)} D(-l^2)]^2 &= \frac{9\beta_0^2 \mu_0^2}{8\pi} \ , \end{aligned} \tag{8}$$

where $\beta_0 \approx 2.0\,\mathrm{GeV}^{-1}$ is the coupling of the pomeron to $u$ and $d$-quarks [16], and $\alpha_s^{(0)}$ the strong coupling constant between on-shell quarks and a low-$l^2$ gluon, which we will take to be $\alpha_s^{(0)} \approx 1$ (cf. [5]). The parameter $\mu_0 \approx 1.1\,\mathrm{GeV}$ has been estimated from exclusive $\rho$ production in deep inelastic scattering [4], and sets the characteristic scale over which $D(l^2)$ decreases. To the results obtained with this propagator one has to add the contribution from the perturbative piece which was subtracted at large gluon momenta. In diffractive hadron-hadron scattering it gives only a small correction [4].

Exchange of two noninteracting gluons is only the crudest approximation to the pomeron. We will calculate the leading term in $\xi^{-1}$ at one-loop level, and in order to make contact with experiment we shall modify by hand the exponent of $\xi$ using the pomeron trajectory

$$\alpha_{I\!P}(t) = 1 + \epsilon + \alpha' t \tag{9}$$

with $\epsilon \approx 0.08$ [17] and $\alpha' \approx 0.25\,\mathrm{GeV}^{-2}$ [16].

The scattering of physical hadrons is calculated from the scattering of their constituent quarks. LN have found that, if the length $\mu_0^{-1}$ is small compared with the hadron radius, the diagrams where the two gluons of the pomeron couple to different quarks in a hadron are suppressed, which naturally leads to the additive quark rule. Let us see what the remaining diagrams give for the radiation of a pomeron off a proton.

We assume that the momentum of a constituent quark in the proton can be approximated by $\tilde{p} \approx xp$ with $x \gg \xi$, neglecting the transverse momentum of the constituents within the fast proton as we have previously neglected the transverse momentum $\Delta_T$ of the scattered proton. The product of Dirac matrices corresponding to the upper quark line in fig. 2 is

$$t_{\beta\alpha} = (\tilde{\not p} - \not\Delta + M_q)\, \gamma_\beta \, (\tilde{\not p} - \not\Delta + \not l + M_q)\, \gamma_\alpha \, (\tilde{\not p} + M_q) \ , \tag{10}$$

where $M_q$ is an effective mass for the constituent quark. Its precise value turns out to be irrelevant for our calculation, since corresponding terms are nonleading in $\xi^{-1}$. Rewriting $\tilde{p} \approx xp \approx x\xi^{-1}\Delta$ we see that to leading order in $\xi^{-1}$

$$\begin{aligned} t_{\beta\alpha} &\approx (\tilde{\not p} + M_q)\, \gamma_\beta \, (\tilde{\not p} + M_q)\, \gamma_\alpha \, (\tilde{\not p} + M_q) \\ &\approx x\xi^{-1}\, \Delta_\beta \, (\tilde{\not p} + M_q)\, \gamma_\alpha \, (\tilde{\not p} + M_q) + (\alpha \leftrightarrow \beta) \ , \end{aligned} \tag{11}$$

[1] The definitions in [4] differ from the ones used here by a colour factor of 2/9 (cf. next section), because the gluons are treated as abelian there.

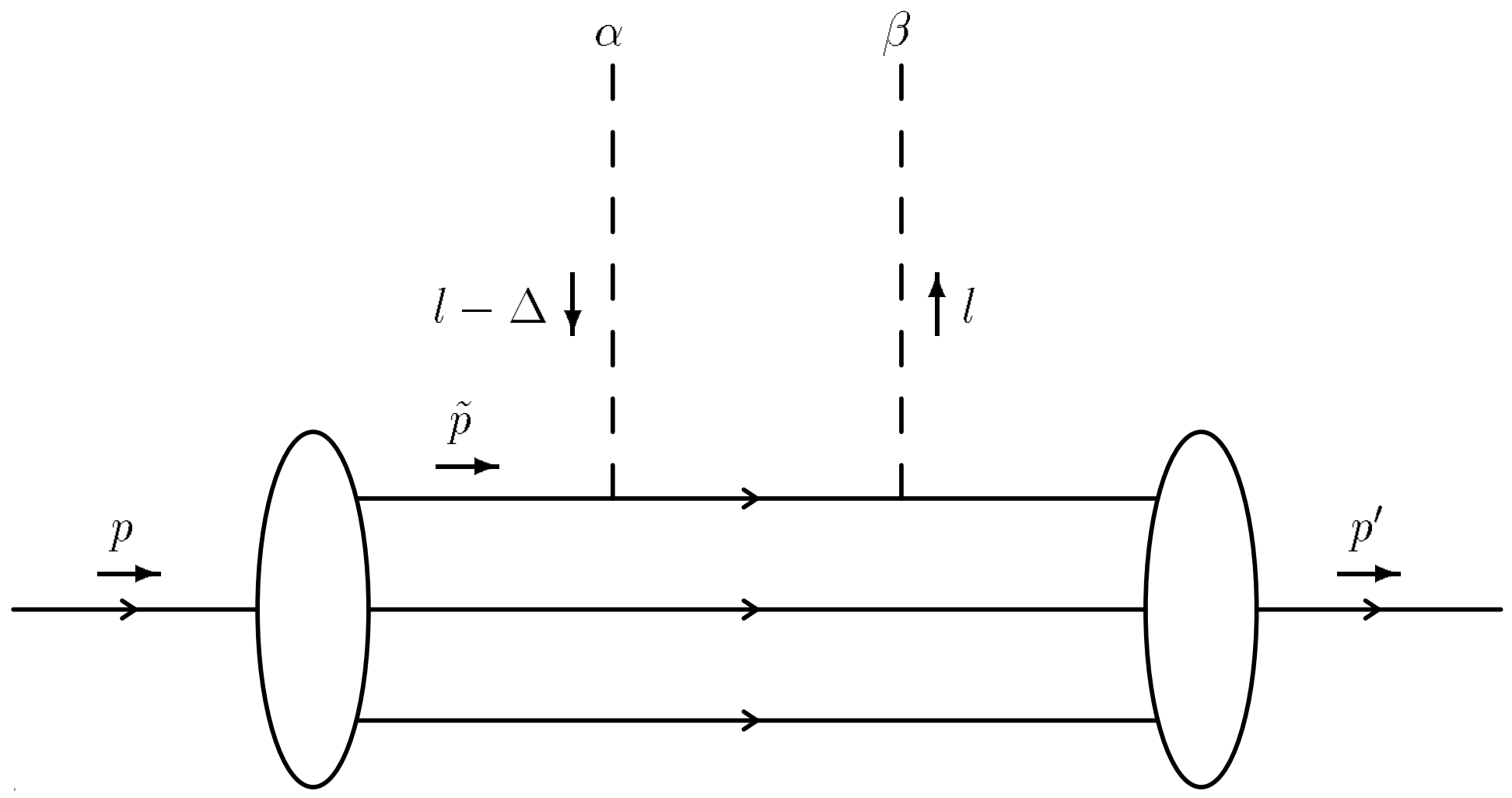

Figure 2: A typical diagram contributing to the interaction between a proton and the pomeron, modelled by two nonperturbative gluons. The blobs stand for to the proton wave function and the dashed lines for gluons with Lorentz indices $\alpha$ and $\beta$.

where in the last step we have used the anticommutation relations for Dirac matrices. The matrix structure of this expression involves just the vector current of the proton, i.e. to leading order the spin structure of the pomeron-proton coupling is the same as for the coupling of the proton to a photon.[2] Summing over the constituent quarks in the scattering amplitude one has for the proton vector current [18]

$$\langle p'|\sum_{q=u,d}\bar{q}(0)\gamma_\alpha q(0)|\,p\rangle \approx 3F_1(t)\,\bar{u}(p')\gamma_\alpha u(p) \;, \tag{12}$$

where the $t$-dependence is controlled by the isoscalar form factor of the nucleon, which we approximate by the Dirac form factor of the proton[3]

$$F_1(t) = \frac{4M_p^2 - 2.8\,t}{4M_p^2 - t}\left(1 - \frac{t}{0.7\,\mathrm{GeV}^2}\right)^{-2} \;. \tag{13}$$

We will find in the next section that the factor $x\xi^{-1}$ in (11) is cancelled by the integration element of the loop momentum $l$, so that the coupling of the proton to the two gluons in the pomeron can be expressed in terms of $\Delta$ and the current in (12).

[2]Note, however, that due to its signature factor the pomeron has positive parity under charge conjugation, unlike the photon.

[3]Since $F_1(t)$ depends strongly on the squared momentum transfer $t \approx \Delta_T^2$ we do not neglect the transverse momentum $\Delta_T$ of the pomeron at this point. The above result (11) remains valid in this case if one replaces $\Delta$ with $\Delta - \Delta_T$

## 4 Calculation of the cross section

We shall now give some details on the calculation of the process $p+\gamma \to p+q\bar{q}$ or $p+\gamma^* \to p+q\bar{q}$ in the LN model.

The leading term in $\xi^{-1}$ comes from the antihermitian part of the transition matrix, which is equal to the imaginary part of the amplitude if an appropriate phase convention for spinors is used,[4] and can be calculated using the unitarity relation for the scattering matrix. Fig. 3 shows the four corresponding Feynman diagrams. The lower fermion line stands for a constituent quark of the proton, which we shall treat as on-shell before and after the scattering.

To project out the colour singlet part of the two-gluon amplitude, we note that the colour structure of all diagrams is given by

$$t^a t^b \otimes t^a t^b = \frac{2}{9} \mathbb{1} \otimes \mathbb{1} - \frac{1}{3} t^c \otimes t^c \ , \tag{14}$$

$t^a$ denoting the Gell-Mann matrices divided by 2. Hence the appropriate factor for the colour-singlet amplitude is 2/9.

For diagrams $(c)$ and $(d)$ we parametrise

$$l = l_T + \eta\Delta + \eta'\tilde{q} \ , \tag{15}$$

where

$$\tilde{q} = q + \frac{Q^2}{\hat{s}+Q^2}\Delta \tag{16}$$

is lightlike up to $O(\xi^2)$ and $l_T$ is transverse to $\Delta$ and $q$. Putting the four quarks coupled to the gluon with momentum $l$ on shell gives a set of quadratic equations in $\eta$ and $\eta'$. We introduce $\hat{t} = (q-P)^2$ and $\hat{u} = (q-P')^2$ as Mandelstam variables of the subreaction $I\!\!P + \gamma \to q\bar{q}$, where $I\!\!P$ denotes the pomeron. The integration element then is, to leading order in $\xi^{-1}$,

$$d^4 l\, \delta\left((P-l)^2 - m^2\right)\, \delta\left((\tilde{p}-\Delta+l)^2 - M_q^2\right) = \frac{d^2 l_T}{2x\xi^{-1}(\hat{u}-m^2)} \ . \tag{17}$$

One solution of the constraints gives, again to leading order,

$$\eta = \frac{\mathbf{l}_T^2 - 2\mathbf{l}_T{\cdot}\mathbf{p}_T}{\hat{u}-m^2} \ , \qquad \eta' = O(\xi) \ , \qquad l^2 = -\mathbf{l}_T^2 \ , \tag{18}$$

[4]This result has been obtained in an approach where the exchanged gluons are treated perturbatively [8]. For a nonperturbative gluon propagator it can be reproduced by a calculation along the lines of [3], where only analyticity properties of $D(l^2)$ are needed.

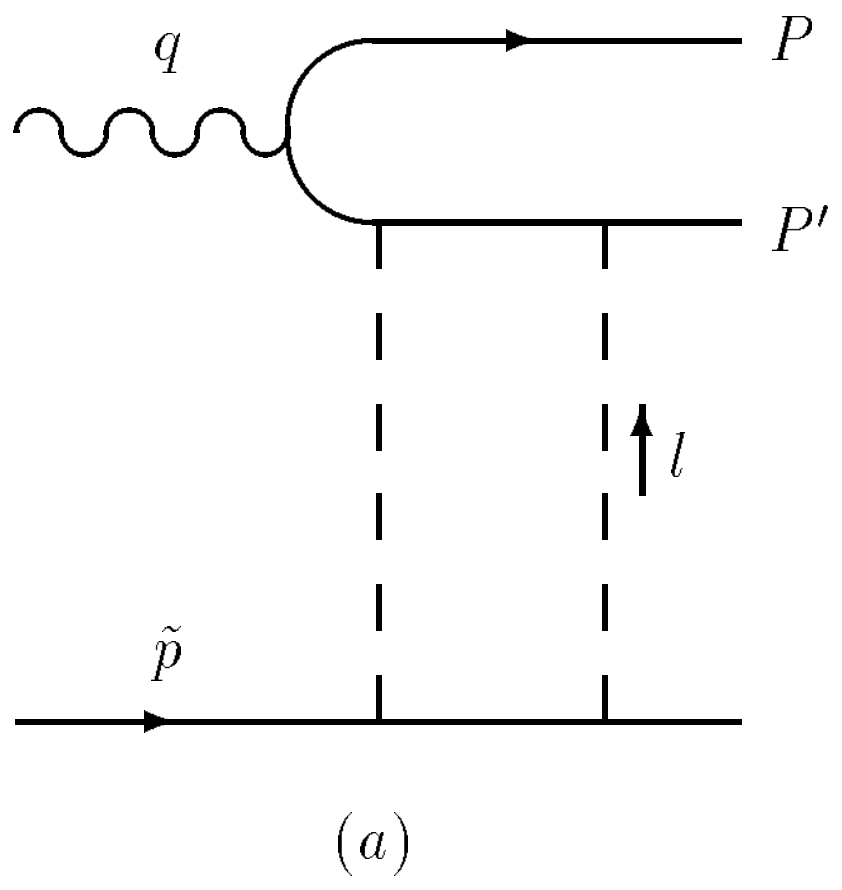

Figure 3: Feynman diagrams contributing to the imaginary part of the amplitude for $p+\gamma \to p+q\bar{q}$. The pomeron is approximated by two nonperturbative gluons (dashed lines).

whereas the other leads to

$$l^2 = x\xi^{-1}(\hat{u} - m^2) \ , \tag{19}$$

which is very large due to the factor $\xi^{-1}$ and therefore has to be discarded. For the other two diagrams we make a similar ansatz

$$l = -l_T + \eta\Delta + \eta'\tilde{q} \tag{20}$$

and, choosing the sign of $l_T$ in this particular way, obtain analogous results with $\hat{u}$ replaced by $\hat{t}$ and $P$ by $P'$ in equations (17) to (19). For the denominator of the off-shell quark propagator we then find $d_a = \hat{t} - m^2$, $d_c = \hat{u} - m^2$ and

$$\frac{d_b}{\hat{u} - m^2} = \frac{d_d}{\hat{t} - m^2} = \frac{1}{\mathbf{p}_T^2 + m^2}\,\frac{\hat{s}}{\hat{s} + Q^2}\left[(\mathbf{l}_T - \mathbf{p}_T)^2 + \frac{Q^2}{\hat{s}}\,(\mathbf{p}_T^2 + m^2) + m^2\right] \tag{21}$$

for diagrams $(a)$, $(c)$, $(b)$ and $(d)$, respectively. The kinematics of the first two of these diagrams are clearly different from those of the other two, and the contributions from the two separate sets will be discussed is some detail below.

Together with (11) and (12) for the coupling of the gluons to the proton it is now straightforward to calculate the differential cross section for our process.[5] We shall give some intermediate results which will be useful in the discussion.

Let $\alpha$ be the electromagnetic fine structure constant and $e_q$ the charge of the produced quarks in units of the elementary charge. In each diagram of fig. 3 there is one quark-gluon vertex which involves an off-shell quark. We take the running strong coupling $\alpha_s(\mathbf{p}_T^2)$ instead of $\alpha_s^{(0)}$ for this vertex in all four diagrams, with the usual uncertainty about which argument should be chosen for the running coupling to keep higher loop corrections small. In our numerical studies we use a fixed $\alpha_s(\mathbf{p}_T^2) = 0.2$, given that the variation of the running coupling is not very large in the relevant range of $\mathbf{p}_T^2$. We use a flux factor $2(W^2 + Q^2) = 2\xi^{-1}(\hat{s} + Q^2)$ for the photon, take the average over the spin of the incoming proton, and sum over the spins of the final state fermions as well as the colour of the produced $q\bar{q}$ pair. Then

$$\frac{d\sigma_{T,L}}{dt\, d\xi\, d\mathbf{p}_T^2} = \frac{16}{3}\,\frac{\alpha_s(\mathbf{p}_T^2)}{\alpha_s^{(0)}}\alpha e_q^2\, \xi^{1-2\alpha_{I\!P}(t)}\, [F_1(t)]^2 \frac{1}{\hat{s} + Q^2}\,\frac{1}{\sqrt{1 - 4(\mathbf{p}_T^2 + m^2)/\hat{s}}}\, \mathcal{I}_{T,L} \ , \tag{22}$$

[5]We have used the algebraic programs FORM by J Vermaseren and MATHEMATICA by Wolfram Research, Inc.

where

$$\mathcal{I}_T = \left(1 - 2\,\frac{\mathbf{p}_T^2 + m^2}{\hat{s}}\right) \frac{1}{\mathbf{p}_T^2 + m^2} \times$$
$$\left(\int \frac{d^2 l_T}{\pi}\, [\alpha_s^{(0)} D(-\mathbf{l}_T^2)]^2 \left[\mathbf{p}_T + (\mathbf{l}_T - \mathbf{p}_T)\frac{\hat{u} - m^2}{d_b}\right]\right)^2$$
$$+ \frac{m^2}{\mathbf{p}_T^2 + m^2} \left(\int \frac{d^2 l_T}{\pi}\, [\alpha_s^{(0)} D(-\mathbf{l}_T^2)]^2 \left[1 - \frac{\hat{u} - m^2}{d_b}\right]\right)^2 \tag{23}$$

and

$$\mathcal{I}_L = 4\,\frac{Q^2}{\hat{s}}\,\frac{\mathbf{p}_T^2 + m^2}{\hat{s}} \left(\int \frac{d^2 l_T}{\pi}\, [\alpha_s^{(0)} D(-\mathbf{l}_T^2)]^2 \times\right.$$
$$\left.\left[\left\{1 - \frac{\hat{s} + Q^2}{2Q^2}\right\} - \left\{\frac{\hat{u} - m^2}{d_b} - \frac{\hat{s} + Q^2}{2Q^2}\right\}\right]\right)^2 \tag{24}$$

for photons with transverse or longitudinal polarisation in the proton-photon centre of mass, respectively. The cross section is the same for the two possible transverse polarisations because we have integrated over the azimuthal angle of the jets.

Note that it is possible to write the longitudinal cross section as the square of an integral over one loop momentum.[6] Similarly, the transverse cross section is the sum of two squared integrals, the integrand in the first term being a vector quantity. By calculating separately the squared amplitudes for the sums of diagrams $(a)$, $(c)$ and $(b)$, $(d)$, as well as their interference term, it is possible to separate the contributions of these two sets to the cross section. In (23) the first term in each of the large brackets comes from the sum of $(a)$ and $(c)$, the second (with $d_b$ in the denominator) from the sum of diagrams $(b)$ and $(d)$. Likewise, in (24), the first expression enclosed in curly braces comes from $(a)$ and $(c)$, the second one from $(b)$ and $(d)$. For the latter, the denominator $d_b$ has cancelled in the term proportional to $1/Q^2$. One can see that the two sets of diagrams are not separately invariant under the electromagnetic gauge group, as the amplitudes for the sums of $(a)$ and $(c)$ and of $(b)$ and $(d)$ diverge for $Q^2 \to 0$. Similarly, it turns out that these sums are nonzero if the photon polarisation is set proportional to its momentum $q$.

One can easily perform the integrations over the azimuthal angle of $l_T$ and then has

[6] It cannot readily be identified as proportional to the amplitude because the cross section is summed over polarisations and azimuthal angles in the final state.

$$\begin{aligned}
\mathcal{I}_T &= \left(1 - 2\,\frac{\mathbf{p}_T^2 + m^2}{\hat{s}}\right)\,\frac{\mathbf{p}_T^2}{\mathbf{p}_T^2 + m^2}\left(\int_0^\infty d\mathbf{l}_T^2\,[\alpha_s^{(0)} D(-\mathbf{l}_T^2)]^2 f_1(v,w)\right)^2 \\
&\quad + \frac{m^2}{\mathbf{p}_T^2 + m^2}\left(\int_0^\infty d\mathbf{l}_T^2\,[\alpha_s^{(0)} D(-\mathbf{l}_T^2)]^2 f_2(v,w)\right)^2 \\
\mathcal{I}_L &= 4\,\frac{Q^2}{\hat{s}}\,\frac{\mathbf{p}_T^2 + m^2}{\hat{s}}\left(\int_0^\infty d\mathbf{l}_T^2\,[\alpha_s^{(0)} D(-\mathbf{l}_T^2)]^2 f_2(v,w)\right)^2 \quad , \qquad (25)
\end{aligned}$$

where we have introduced the functions

$$\begin{aligned}
f_1(v,w) &= 1 + \frac{w}{2}\left[\frac{v + w - 2}{\sqrt{(v+w-2)^2 - 4(1-w)}} - 1\right] \\
f_2(v,w) &= 1 - \frac{w}{\sqrt{(v+w-2)^2 - 4(1-w)}} \qquad (26)
\end{aligned}$$

of the arguments

$$v = \frac{\mathbf{l}_T^2}{\mathbf{p}_T^2}\ , \qquad w = \frac{\mathbf{p}_T^2 + m^2}{\mathbf{p}_T^2}\,\frac{\hat{s} + Q^2}{\hat{s}}\ . \qquad (27)$$

Fig. 4 shows $f_1(v,w)$ and $f_2(v,w)$ for some fixed values of $w$. Independent of $w$, both $f_1$ and $f_2$ vanish at $v = 0$ and tend to 1 as $v \to \infty$. For $w = 1$, $f_1$ becomes the step function $\theta(v-1)$. The singularity of $f_2$ at $v = 1$ in the same limit is spurious, since $w$ only becomes 1 if both $Q^2$ and $m^2$ go to zero, and the integrals with $f_2$ are multiplied with one of these factors.

This representation of the cross section is similar to the one obtained by Nikolaev and Zakharov [9], who work with a perturbative gluon propagator. Our equations (22) and (25) to (27), taken at $t = 0$, are in agreement with eq. (22) to (27) of ref. [9][7] if

1. we replace $\alpha_{\mathbb{P}}(t)$ in our eq. (22) with 1, i. e. if we do not introduce the pomeron trajectory in order to account for the interaction between the gluons,

2. one takes into account that the strong coupling is taken at different momentum scales in the present paper and in [9],

3. our $[D(-\mathbf{l}_T^2)]^2$ is replaced with $V(\mathbf{l}_T)/(\mathbf{l}_T^2 + \mu_G^2)^2$, corresponding to the different assumptions in [9] about the gluon propagator and the coupling between the gluons in the pomeron to a hadron,

[7] We assume that in the second line of their eq. (24) it should read $\boldsymbol{\kappa}^2$ instead of $\boldsymbol{\kappa}$.

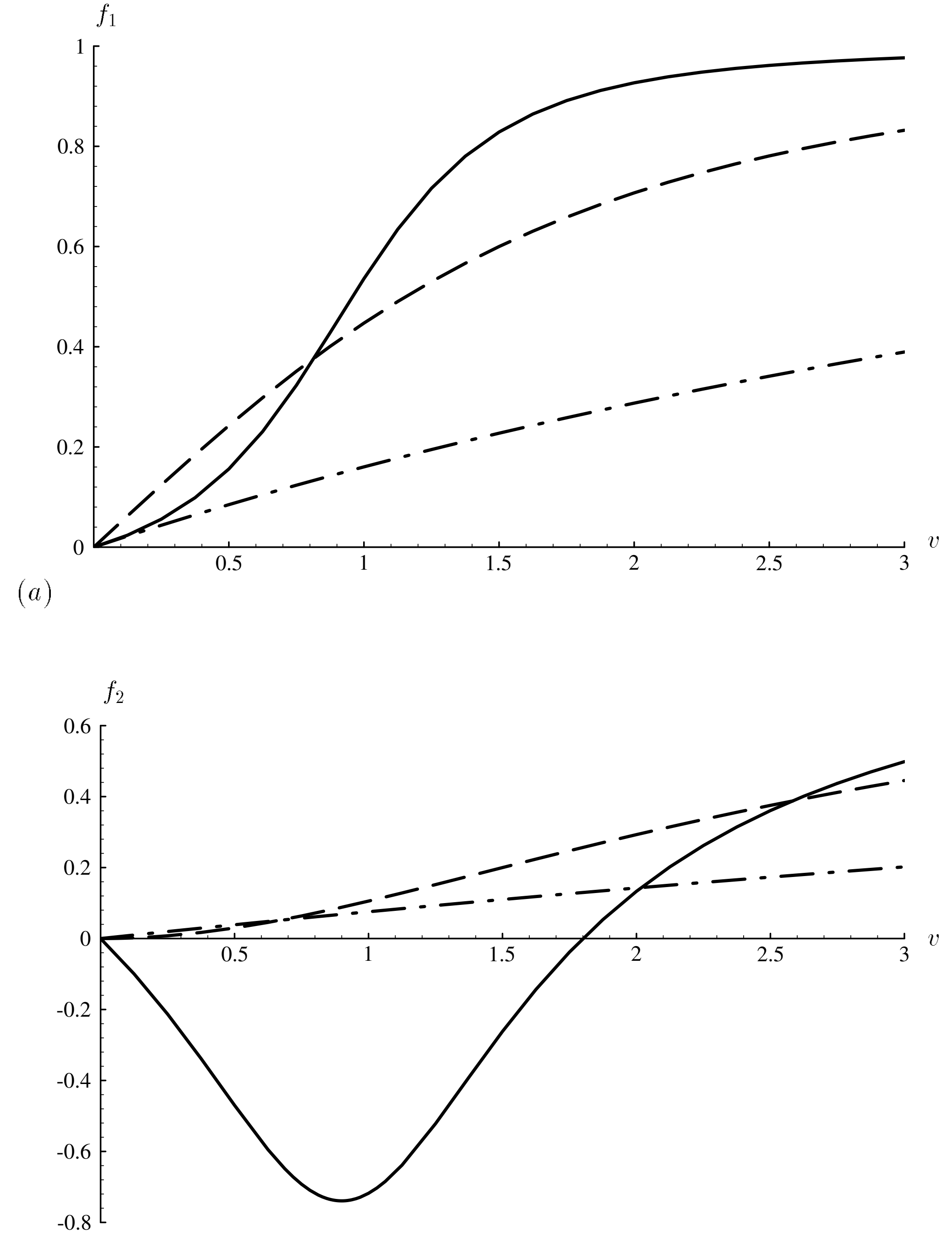

Figure 4: The functions $f_1$ and $f_2$ of eq. (26) for $w = 1.1$ (full), $w = 2$ (dashed) and $w = 10$ (dot-dashed). $f_2$ has a zero at $v = 2(2 - w)$.

4. one makes a transformation from our integration variable $\xi$ to their $\alpha$ with
$$\hat{s} = \frac{\mathbf{p}_T^2 + m^2}{\alpha(1-\alpha)} \tag{28}$$
(eq. (19) of [9]). Using the relation (5) between $\xi$ and $\hat{s}$ we find
$$\begin{aligned} d\alpha &= 2\, d\hat{s}\, \frac{\mathbf{p}_T^2 + m^2}{\hat{s}^2} \left(1 - 4\, \frac{\mathbf{p}_T^2 + m^2}{\hat{s}}\right)^{-1/2} \\ &= 2\, d\xi\, \xi^{-1}\, \frac{\hat{s} + Q^2}{\hat{s}}\, \frac{\mathbf{p}_T^2 + m^2}{\hat{s}} \left(1 - 4\, \frac{\mathbf{p}_T^2 + m^2}{\hat{s}}\right)^{-1/2} . \end{aligned} \tag{29}$$
Note that for fixed $\mathbf{p}_T^2$ and $\hat{s}$ there are two values of $\alpha$, which gives a factor of 2 in the integration element as explained at the end of section 2.[8]

Assuming that due to the squared gluon propagator $[D(-\mathbf{l}_T^2)]^2$ the main contribution to the loop integrals comes from values of $\mathbf{l}_T^2$ up to the order of $\mu_0^2 \approx 1.2\,\mathrm{GeV}^2$, which is small compared to the minimum $\mathbf{p}_T^2$ required for the jets, we can expand $f_1$ and $f_2$ to first order in $\mathbf{l}_T^2/\mathbf{p}_T^2$ and obtain, using the second moment in (8),
$$\begin{aligned} \mathcal{I}_T &\approx \frac{81 \beta_0^4 \mu_0^4}{16\pi^2}\, \frac{\hat{s}^2}{(\hat{s}+Q^2)^2}\, \frac{1}{(\mathbf{p}_T^2 + m^2)^2} \\ &\quad \left\{ \left(1 - 2\, \frac{\mathbf{p}_T^2 + m^2}{\hat{s}}\right) \frac{\mathbf{p}_T^2}{\mathbf{p}_T^2 + m^2} \left(1 - \frac{\hat{s}}{\hat{s}+Q^2}\, \frac{\mathbf{p}_T^2}{\mathbf{p}_T^2 + m^2}\right)^2 \right. \\ &\quad \left. + \frac{1}{4}\, \frac{m^2}{\mathbf{p}_T^2 + m^2} \left(1 - 2\, \frac{\hat{s}}{\hat{s}+Q^2}\, \frac{\mathbf{p}_T^2}{\mathbf{p}_T^2 + m^2}\right)^2 \right\} \\ \mathcal{I}_L &\approx \frac{81 \beta_0^4 \mu_0^4}{16\pi^2}\, \frac{Q^2}{(\hat{s}+Q^2)^2}\, \frac{1}{\mathbf{p}_T^2 + m^2} \left(1 - 2\, \frac{\hat{s}}{\hat{s}+Q^2}\, \frac{\mathbf{p}_T^2}{\mathbf{p}_T^2 + m^2}\right)^2 . \end{aligned} \tag{30}$$
To assess how good this approximation is, we have evaluated the loop integrals in (25) for a specific model of the gluon propagator, namely [4]
$$D(-\mathbf{l}_T^2) \propto \left(1 + \frac{\mathbf{l}_T^2}{(n-1)\mu_0^2}\right)^{-n} , \tag{31}$$
where $n \geq 4$ and the proportionality constant can easily be obtained from (8). For $n \to \infty$ this becomes $D(-\mathbf{l}_T^2) \propto \exp(-\mathbf{l}_T^2/\mu_0^2)$. The approximation in (30) improves slightly with increasing $n$. Fig. 5 shows that it is rather good, although at small $Q^2/\hat{s}$ it underestimates the cross section for transverse photons.[9]

[8] We differ from eq. (6) of [11] by this factor.

[9] If at $Q^2 = 5\,\mathrm{GeV}^2$, $\hat{s} = 400\,\mathrm{GeV}^2$ and $m = 0$, we integrate the transverse differential

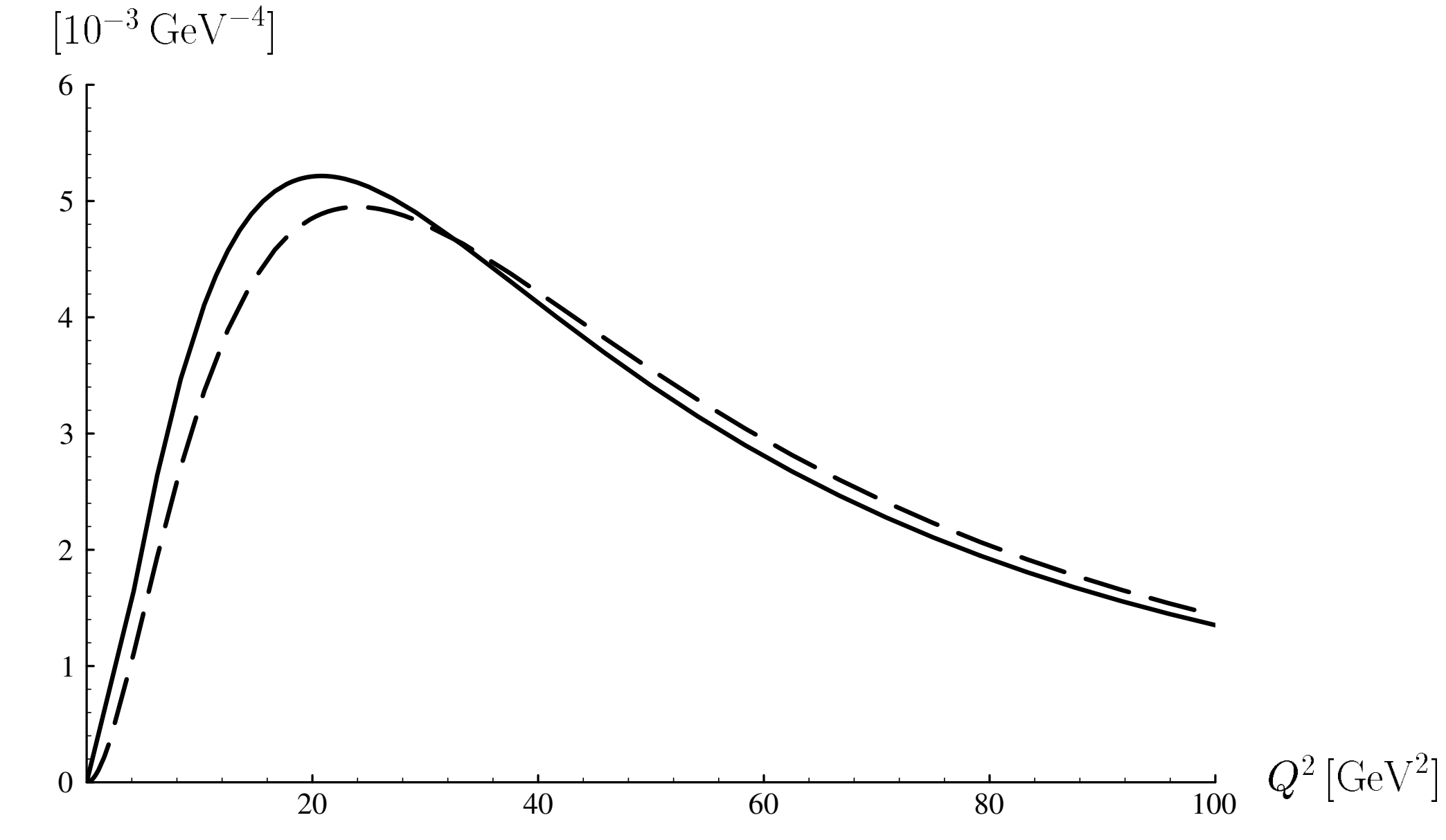

(a)

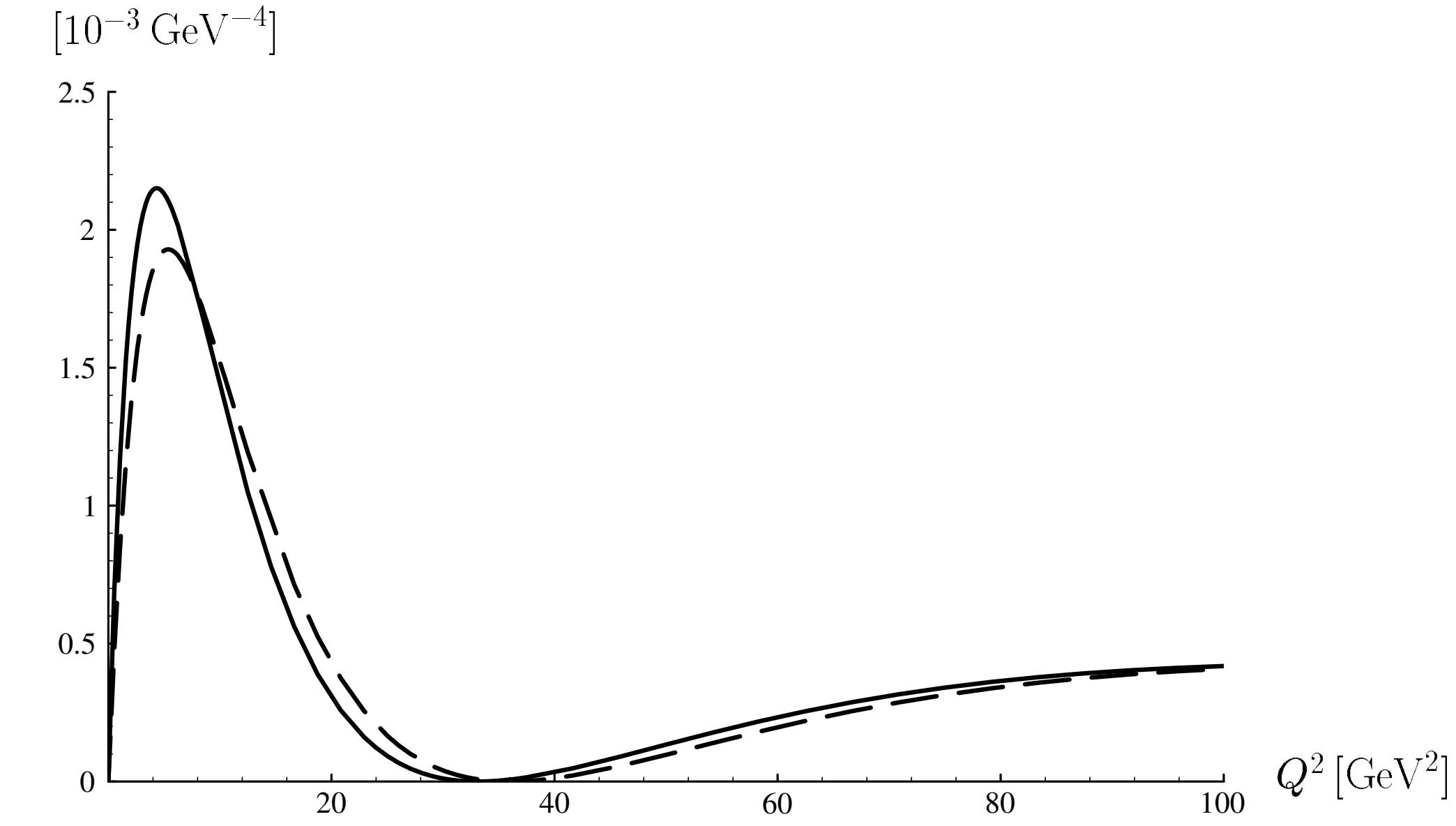

(b)

Figure 5: ($a$) The squared loop integral of eq. (25) which involves $f_1$ (full), compared with its approximation in (30) (dashed) for $|\mathbf{p}_T| = 3\,\mathrm{GeV}$, $\hat{s} = 4\mathbf{p}_T^2$, $m = 0$ and $n = 4$. To obtain the correct $Q^2$-dependence of the cross section it has been multiplied with $\hat{s}/(\hat{s} + Q^2)$. Up to this global factor, changing $\hat{s}$ and $m$ corresponds to rescaling $Q^2$ since $f_1$ depends on these parameters only through $w$ in (27). ($b$) Same as ($a$) but for $f_2$ instead of $f_1$ and with a global factor of $Q^2/(\hat{s}+Q^2)$ instead of $\hat{s}/(\hat{s}+Q^2)$, so as to obtain the $Q^2$-behaviour of the longitudinal cross section.

In the calculation of ref. [6] the propagator denominators $d_b$ and $d_d$, which depend on the loop momentum, have been approximated by their values at $\mathbf{l}_T = 0$. This leads to a wrong result, because at $\mathbf{l}_T^2 = 0$ the integrands vanish, and to obtain the terms of order $\mathbf{l}_T^2$ the propagators have to be expanded to this order, too.

# 5 Results for photoproduction

We will now discuss our results for the case of photoproduction, concentrating on the production of light quarks. Setting $m = 0$ in eq. (30) one finds that the cross section for transverse photons is proportional to $Q^4$, and thus vanishes for photoproduction even faster than the longitudinal one which behaves like $Q^2$. This is not an artifact of the linear approximation in $\mathbf{l}_T^2/\mathbf{p}_T^2$ which led to (30). At $m = 0$ and $Q^2 = 0$ the function $f_1$ in (25) is the step function $\theta(\mathbf{l}_T^2 - \mathbf{p}_T^2)$ and the loop integral is strongly suppressed by a factor of order $[D(-\mathbf{p}_T^2)]^2$. For squared momenta of order $\mathbf{p}_T^2$ the *perturbative* part of the gluon propagator is of course dominant. To calculate its contribution one can substitute $1/(-\mathbf{l}_T^2)$ for $D(-\mathbf{l}_T^2)$ and $\alpha_s(\mathbf{p}_T^2)$ for $\alpha_s^{(0)}$, taking all four quark-gluon vertices in fig. 3 as perturbative and using the perturbative coupling at the fixed scale $\mathbf{p}_T^2$ for simplicity. Then the cross section for $m = 0$ and $Q^2 = 0$ becomes (cf. (22) and (25))

$$\frac{d\sigma}{dt\, d\xi\, d\mathbf{p}_T^2} = \frac{16}{3}[\alpha_s(\mathbf{p}_T^2)]^4 \alpha e_q^2\, \xi^{1-2\alpha_{I\!P}(t)} [F_1(t)]^2 \, \frac{1}{\hat{s}} \frac{1}{\sqrt{1 - 4\mathbf{p}_T^2/\hat{s}}} \left(1 - 2\,\frac{\mathbf{p}_T^2}{\hat{s}}\right) \frac{1}{\mathbf{p}_T^4} \;. \quad (32)$$

With $|\mathbf{p}_T| \geq 3\,\mathrm{GeV}$ and $\xi \leq 0.01$ this gives a very small integrated $\gamma p$ cross section of 9.5 pb for $u$ and $d$-quark jets at a photon-proton centre-of-mass energy of $W = 200\,\mathrm{GeV}$. The effect of imposing an upper limit in the integration over $\xi$ is of the order of 10%.

We must however bear in mind that because of the highly virtual gluons we are no longer dealing with the *soft* pomeron known from phenomenology. The introduction of the soft pomeron trajectory $\alpha_{I\!P}(t)$ to take into account higher order corrections to the exchange of two noninteracting gluons in fig. 3 is presumably no longer justified. We will not attempt to discuss here the difficult problem of the interplay between the soft pomeron and its perturbative counterpart, but only mention that if one substitutes for example the term $\xi^{-2\epsilon}$ in (32) which comes from the soft pomeron trajectory with $\xi^{-1}$ corresponding to the Lipatov value for the pomeron intercept, the integrated

---

cross section over $\mathbf{p}_T^2$ from $9\,\mathrm{GeV}^2$ to its upper kinematical limit, the approximation gives 70% of the exact value for $n = 4$.

cross section mentioned above becomes as large as 1.8 nb.[10] Another question is whether the coupling of the Lipatov pomeron to the proton is dominated by soft or hard scales. In the approximation of two noninteracting gluons, the quark-gluon coupling has of course to be perturbative at the bottom of the diagrams in fig. 3 if it is perturbative at the top.

Let us analyse in some detail what leads to the suppression of loop momenta $\mathbf{l}_T^2 < \mathbf{p}_T^2$ for photoproduction and massless quarks. Eq. (21) for the denominators of the off-shell quark propagators in diagrams $(b)$ and $(d)$ now reads

$$\frac{d_b}{\hat{u}} = \frac{d_d}{\hat{t}} = \frac{(\mathbf{l}_T - \mathbf{p}_T)^2}{\mathbf{p}_T^2} \ , \tag{33}$$

and from eq. (23) we take that the loop integral to be calculated is

$$\int \frac{d^2 l_T}{\pi} \left[D(-\mathbf{l}_T^2)\right]^2 \left[\mathbf{p}_T + (\mathbf{l}_T - \mathbf{p}_T) \frac{\mathbf{p}_T^2}{(\mathbf{l}_T - \mathbf{p}_T)^2}\right] \tag{34}$$

or the analogous expression with $1/\mathbf{l}_T^4$ instead of $[D(-\mathbf{l}_T^2)]^2$. The first term in large brackets comes from diagrams $(a)$ and $(c)$, the second from $(b)$ and $(d)$. The integral for the contribution from $(b)$ and $(d)$ is well known:[11] for $\mathbf{l}_T^2 > \mathbf{p}_T^2$ it averages to zero when the azimuthal angle is integrated out, so that only $(a)$ and $(c)$ contribute to the cross section, whereas for $\mathbf{l}_T^2 < \mathbf{p}_T^2$ it gives a finite contribution which just cancels that from $(a)$ and $(c)$.

Since the squared gluon propagator decreases rapidly with $\mathbf{l}_T^2$ the loop integral is dominated by the region where $\mathbf{l}_T^2$ is not too large compared with $\mathbf{p}_T^2$. In a small part of this region, namely when $\mathbf{l}_T \approx \mathbf{p}_T$, it follows from (33) that in diagrams $(b)$ and $(d)$ the real photon decays into two quarks which are on or near shell and as a consequence have to be approximately collinear with the photon direction. In this case one can expect important strong corrections to the coupling between the photon and the quarks, which in our calculation was taken as the electromagnetic Born-level vertex.

Let us finally investigate the effect of a nonzero quark mass. For light quarks, the contribution to the cross section from the term with $f_1$ in (25) is completely negligible for the same reason as above, since the parameter $w$ in (27) is very close to the value 1, which it takes in the case of zero mass. There is, however, a small contribution from the term proportional to $m^2$ which involves $f_2$. Taking $m = 150\,\mathrm{MeV}$ for strange quarks the nonperturbative contribution to the integrated cross section is very small ($\approx 2\,\mathrm{pb}$). As indicated at the beginning of this section, the transverse cross section calculated with the nonperturbative propagator $D(l^2)$ vanishes faster than the longitudinal one for $Q^2 \to 0$. For production of $u$, $d$ and $s$ quarks at $Q^2 = 1\,\mathrm{GeV}^2$ the former is 3.4 pb whereas the latter is 45 pb, with $W = 200\,\mathrm{GeV}$, $|\mathbf{p}_T| \geq 3\,\mathrm{GeV}$ and $\xi \leq 0.01$.

[10] Here we have used the slope $\alpha'$ of the soft pomeron, but the numerical effect of changing its value is not too great.

[11] In two-dimensional electrostatics it occurs e. g. when the electric field of a rotationally symmetric charge distribution is calculated from Gauss' law.

To see how sensitive our results are to the choice of taking current masses for the produced quarks we have also done the calculation with constituent masses, 300 MeV for $u$ and $d$ and 450 MeV for $s$. Then, due to the term proportional to $m^2$ in (25), the nonperturbative contribution to photoproduction of the three light quark species is 51 pb with $W$, $|\mathbf{p}_T|$ and $\xi$ as above.

In the case of heavy quarks the situation changes and becomes similar to that in DIS. Photoproduction of charm quarks will therefore be treated in the next section.

# 6 Results for deep inelastic scattering

When $Q^2$ is sufficiently large, small $\mathbf{l}_T^2$ are no longer suppressed in $f_1$, because $w$ in (27) is larger than 1. The same is true even at low $Q^2$ for a large quark mass, such as for charm quarks with $m = 1.5\,\mathrm{GeV}$. From

$$\hat{t} - m^2 = -\frac{\hat{s} + Q^2}{2}\left(1 \pm \sqrt{1 - 4(\mathbf{p}_T^2 + m^2)/\hat{s}}\right) \tag{35}$$

and the corresponding equation for $\hat{u}$ with the opposite sign in front of the square root, one can see that the denominators (21) of the propagators in diagrams ($b$) and ($d$) are limited by

$$|d_b|, |d_d| > Q^2\,\frac{\mathbf{p}_T^2 + m^2}{\hat{s}} + m^2 \;\; . \tag{36}$$

Therefore the corresponding quarks are well off-shell and the use of the Born level coupling with the photon, as well as that of the perturbative quark propagator, is justified.

Let us us now discuss some numerical features of the cross section calculated in the LN model. Table 1 gives integrated cross sections obtained from our approximate formulae (30). For light quarks, the longitudinal cross section $\sigma_L$ is larger than the transverse one at very low $Q^2$ (which is still an effect of the low-$\mathbf{l}_T^2$ suppression in $f_1$), but the transverse contribution overtakes between $Q^2 = 10\,\mathrm{GeV}^2$ and $Q^2 = 15\,\mathrm{GeV}^2$. The transverse cross section $\sigma_T$ reaches its largest values around $Q^2 = 40\,\mathrm{GeV}^2$, and becomes quite small beyond $100\,\mathrm{GeV}^2$. For longitudinal photons, the cross section decreases for $Q^2 > 10\,\mathrm{GeV}^2$ and is rather small when $\sigma_T$ takes its maximum. We should remark that the cross sections given in table 1 are lowest-order results which will be modified by perturbative QCD corrections to the upper part of the diagrams in fig. 3, in particular at large values of $Q^2$.

| $Q^2$ [GeV$^2$] | $\sigma_T$[pb] | | | $\sigma_L$[pb] | | |
|---|---|---|---|---|---|---|
| | jet parent quarks | | | | | |
| | $u+d+s$ current masses | $u+d+s$ constituent masses | $c$ | $u+d+s$ current masses | $u+d+s$ constituent masses | $c$ |
| 0 | — | 51 | 300 | — | — | — |
| 5 | 53 | 89 | 270 | 180 | 170 | 33 |
| 10 | 130 | 160 | 250 | 200 | 190 | 39 |
| 15 | 190 | 210 | 240 | 170 | 160 | 35 |
| 20 | 240 | 250 | 240 | 130 | 130 | 30 |
| 40 | 290 | 290 | 200 | 51 | 49 | 14 |
| 60 | 260 | 260 | 160 | 24 | 24 | 8 |
| 80 | 220 | 220 | 130 | 15 | 15 | 7 |
| 100 | 190 | 190 | 110 | 12 | 12 | 6 |
| 150 | 110 | 110 | 58 | 10 | 10 | 6 |
| 200 | 58 | 58 | 29 | 9 | 9 | 6 |
| 250 | 25 | 24 | 12 | 8 | 8 | 5 |

Table 1: Integrated photon-proton cross sections obtained from (30) with a transverse jet momentum $|\mathbf{p}_T| \geq 3\,\mathrm{GeV}$ and a fixed photon-proton centre-of-mass energy of $W = 200\,\mathrm{GeV}$. The effect of the cut $\xi \leq 0.01$ is rather small.

Whether one takes current or constituent masses for the light quarks does not make much difference for the results. The mass dependence is largest at low $Q^2$, where taking constituent masses one obtains a somewhat larger transverse cross section. For $Q^2 \geq 15\,\mathrm{GeV}^2$ the change in $\sigma_T$ when taking current or constituent masses is below 10%, and the same is true for the different values of $\sigma_T$ for $u, d$ or $s$ quarks (apart from the different electric charges). In $\sigma_L$ the effect is yet smaller at all values of $Q^2$.

For charm quarks, the results are quite different. As we have mentioned in the previous section, the transverse cross section does not vanish at $Q^2 = 0$. In fact, it takes its largest value there and decreases with $Q^2$ at fixed $W$. The longitudinal cross section is negligible for all values of $Q^2$.

As for the dependence on the centre-of-mass energy, the cross section rises slowly with $W$. As an example, with $Q^2 = 50\,\mathrm{GeV}^2$, $|\mathbf{p}_T| \geq 3\,\mathrm{GeV}$ and $\xi \leq 0.01$ we obtain a transverse cross section for $u$, $d$ and $s$ production (taking current masses) of 210 pb, 280 pb and 300 pb for $W = 140\,\mathrm{GeV}$, 200 GeV and 250 GeV, respectively. This behaviour can be understood if, using (5), one makes a transformation

$$\frac{d\xi}{\xi} = \frac{d\hat{s}}{\hat{s} + Q^2} \tag{37}$$

and rewrites (22) and (25) in the form

$$d\sigma \propto \xi^{2(1-\alpha_{I\!P}(t))}\,[F_1(t)]^2\,dt\,d\hat{s}\,d\mathbf{p}_T^2 \ , \tag{38}$$

where the constant of proportionality depends on the variables $\hat{s}$, $\mathbf{p}_T^2$ and $Q^2$ only. Expressing $\xi$ in terms of $W^2$ and $\hat{s}$, we see that the only dependence on $W$ comes from the term $\xi^{2(1-\alpha_{I\!P}(t))}$ with its small exponent. The biggest effect is actually that the phase space is reduced at small $W$. We impose an upper limit on $\xi$, and as a consequence there is a maximum allowed $\hat{s}$ increasing with $W$. If we integrate without any upper bound for $\xi$, the above cross sections become 270 pb, 290 pb and 300 pb with an even weaker dependence on $W$. We also see that except for rather low values of $W$ the effect of the cut on $\xi$ is small, for $W = 200\,\text{GeV}$ the results with or without cut differ by less than 4%.

Some features of partially integrated cross section are shown in figures 6 and 7. The $\xi$-spectra in fig. 6 show that for $W = 200\,\text{GeV}$ the cross section is quite small at $\xi = 0.01$. One might remark that the differential cross section for longitudinal photons has a zero for certain values of the parameters. This can be most easily seen from the term in large brackets in (30), which vanishes at $Q^2 = \hat{s}$ for $m = 0$ and at smaller $Q^2$ if $m \neq 0$. Eq. (30) is only an approximation, and the true position of the zero is slightly different (fig. 5). That a zero does exist also follows from the fact that $f_2$ changes its sign at low $\mathbf{l}_T^2$ as $w$ increases (cf. fig. 4), so the loop integral involving $f_2$ changes sign as well and must have a zero for some value of $w$. Due to the rather high values of $Q^2$ at which this occurs it will however be very difficult to see the zero in experiment, because the transverse cross section is comparatively large in this region.

The $\mathbf{p}_T^2$-spectra in fig. 7 are fairly well described by a power behaviour $d\sigma/d\mathbf{p}_T^2 \propto |\mathbf{p}_T|^{-\delta}$. For production of the three light quark flavours with transverse photons, average values of the exponent are $\delta \approx 8.7$ at $Q^2 = 10\,\text{GeV}^2$ and $\delta \approx 6.4$ at $Q^2 = 40\,\text{GeV}^2$, and with longitudinal photons we find $\delta \approx 6$ at $Q^2 = 10\,\text{GeV}^2$. The spectra for charm quarks are less steep, with exponents $\delta \approx 6.7$ at $Q^2 = 10\,\text{GeV}^2$ and $\delta \approx 5.7$ at $Q^2 = 40\,\text{GeV}^2$, both for transverse photon polarisation.

We finally estimate how the contribution from the perturbative gluon propagator at large $\mathbf{l}_T^2$ will modify these results. There is some arbitrariness in choosing the value of $|\mathbf{l}_T|$ above which it should be taken into account [4], but it should be at the GeV-scale, and thus of the same order as the minimum $|\mathbf{p}_T|$ required for the jets. Using a lower cutoff for the $\mathbf{l}_T^2$-integration of $\kappa^2 = 4\,\text{GeV}^2$ and replacing $\alpha_s^{(0)} D(-\mathbf{l}_T^2)$ with $\alpha_s(\mathbf{l}_T^2)/(-\mathbf{l}_T^2)$,[12] we have studied the ratio between the perturbative and the nonperturbative contributions to the differential cross section (22), (30). In the case of transverse

[12] One can also take the running coupling at $\mathbf{p}_T^2$ instead of $\mathbf{l}_T^2$ for the vertex involving the off-shell quark. The numerical difference is very small.

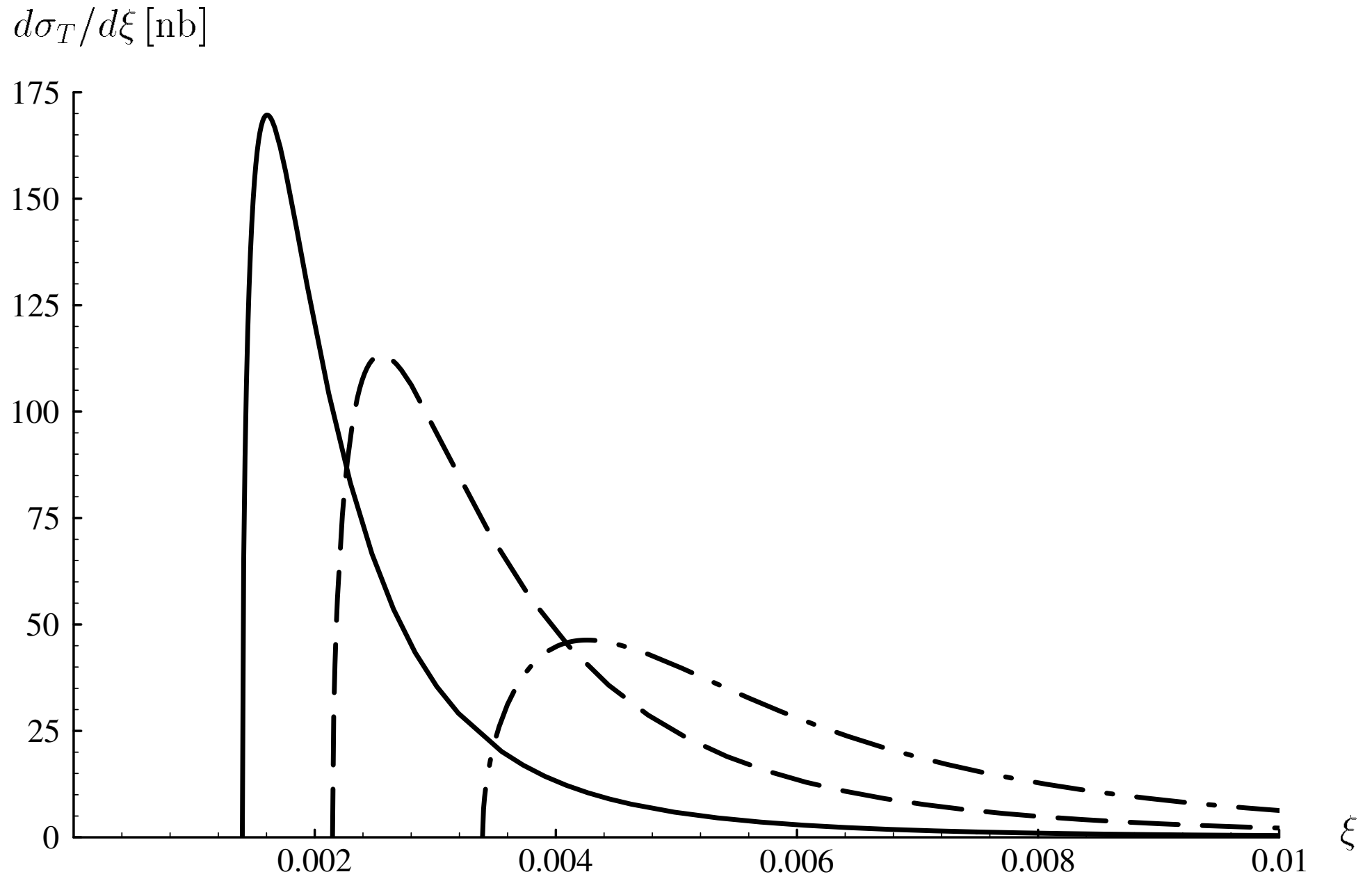

$(a)$

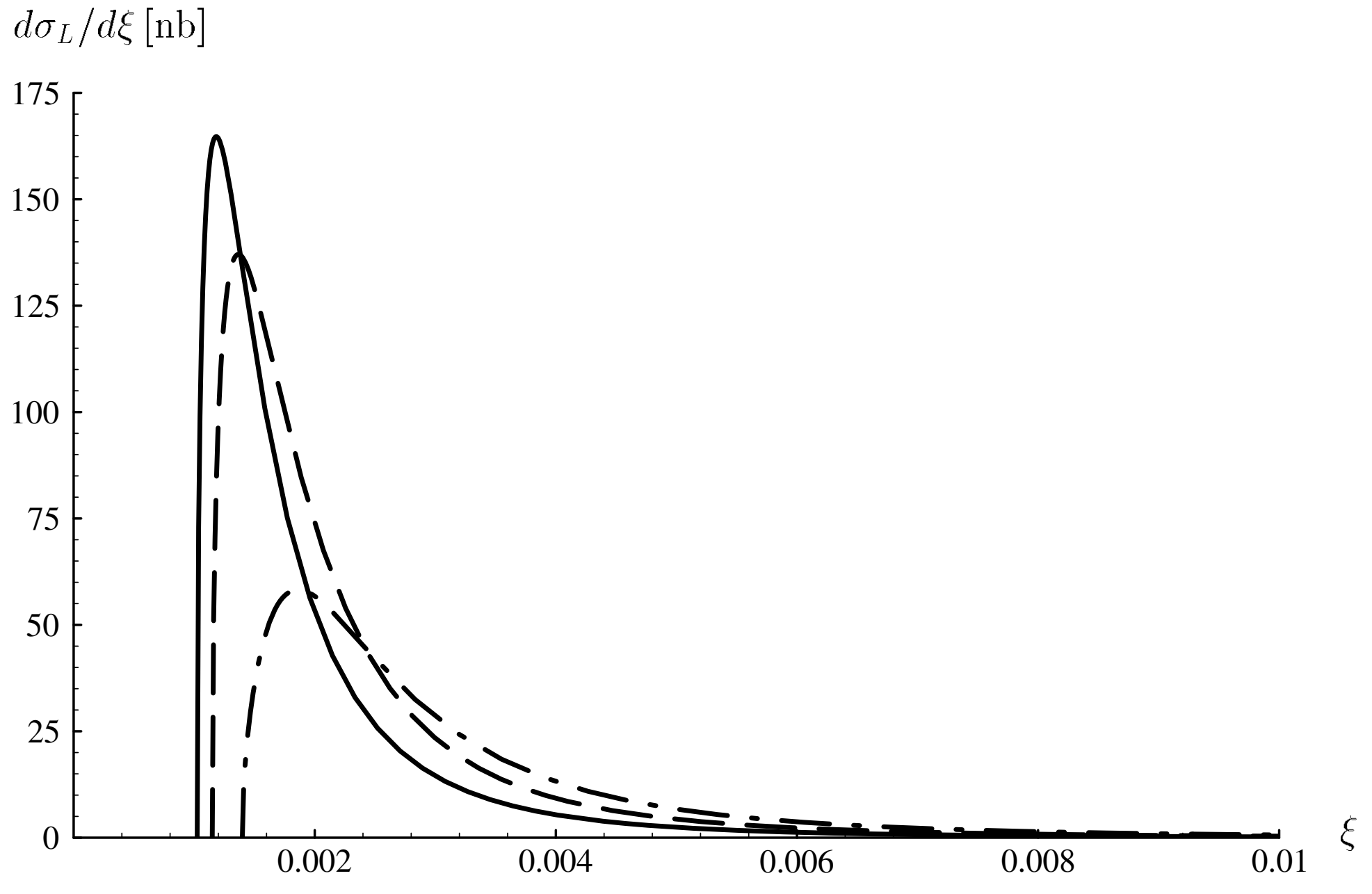

$(b)$

Figure 6: $(a)$ $d\sigma_T/d\xi$ summed over $u$ and $d$ quarks ($m = 0$, $|\mathbf{p}_T| \geq 3\,\mathrm{GeV}$, $W = 200\,\mathrm{GeV}$) at $Q^2 = 20\,\mathrm{GeV}^2$ (full), $50\,\mathrm{GeV}^2$ (dashed), $100\,\mathrm{GeV}^2$ (dot-dashed). $(b)$ The same for $d\sigma_L/d\xi$ at $Q^2 = 5\,\mathrm{GeV}^2$ (full), $10\,\mathrm{GeV}^2$ (dashed), $20\,\mathrm{GeV}^2$ (dot-dashed).

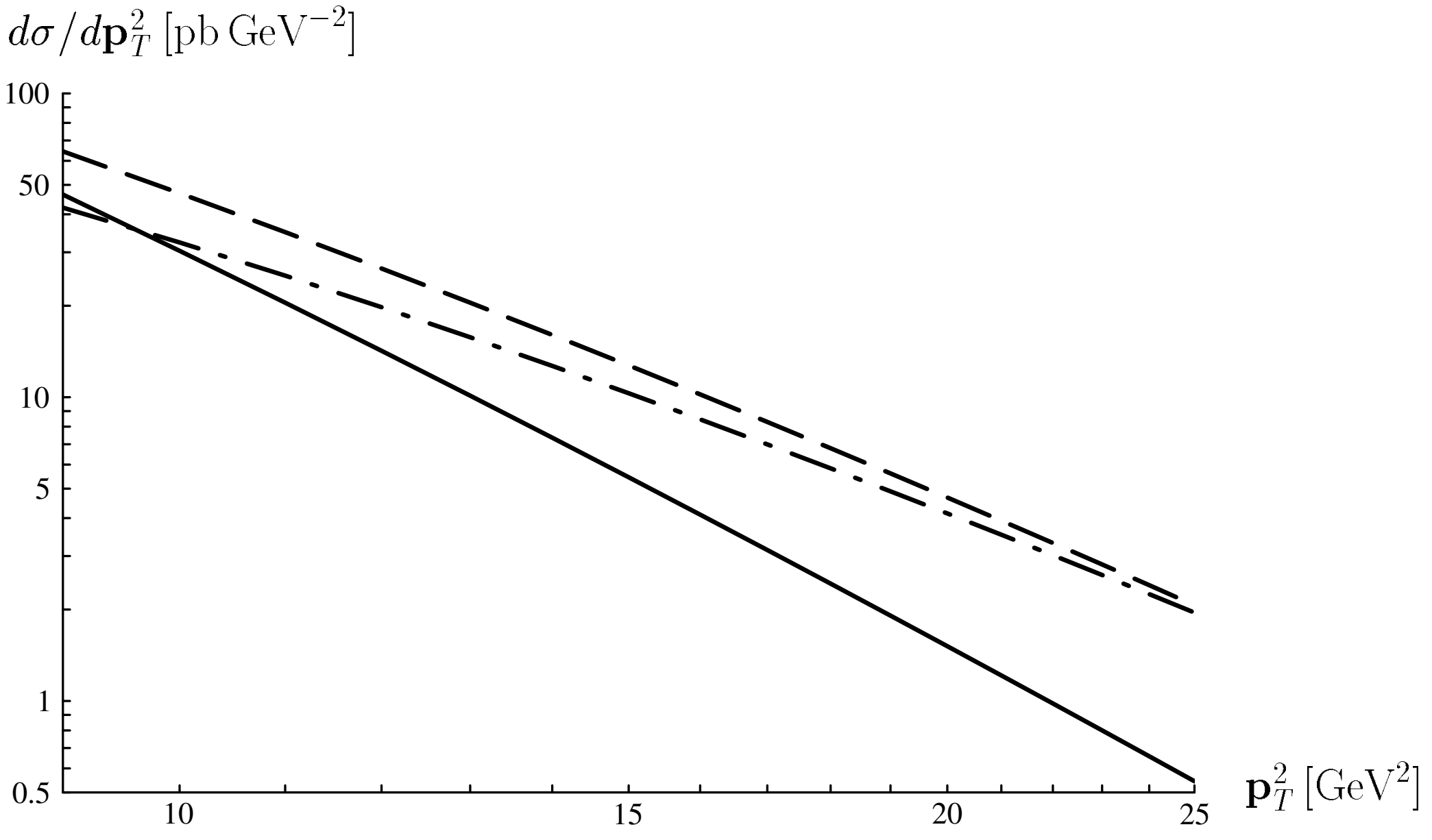

($a$)

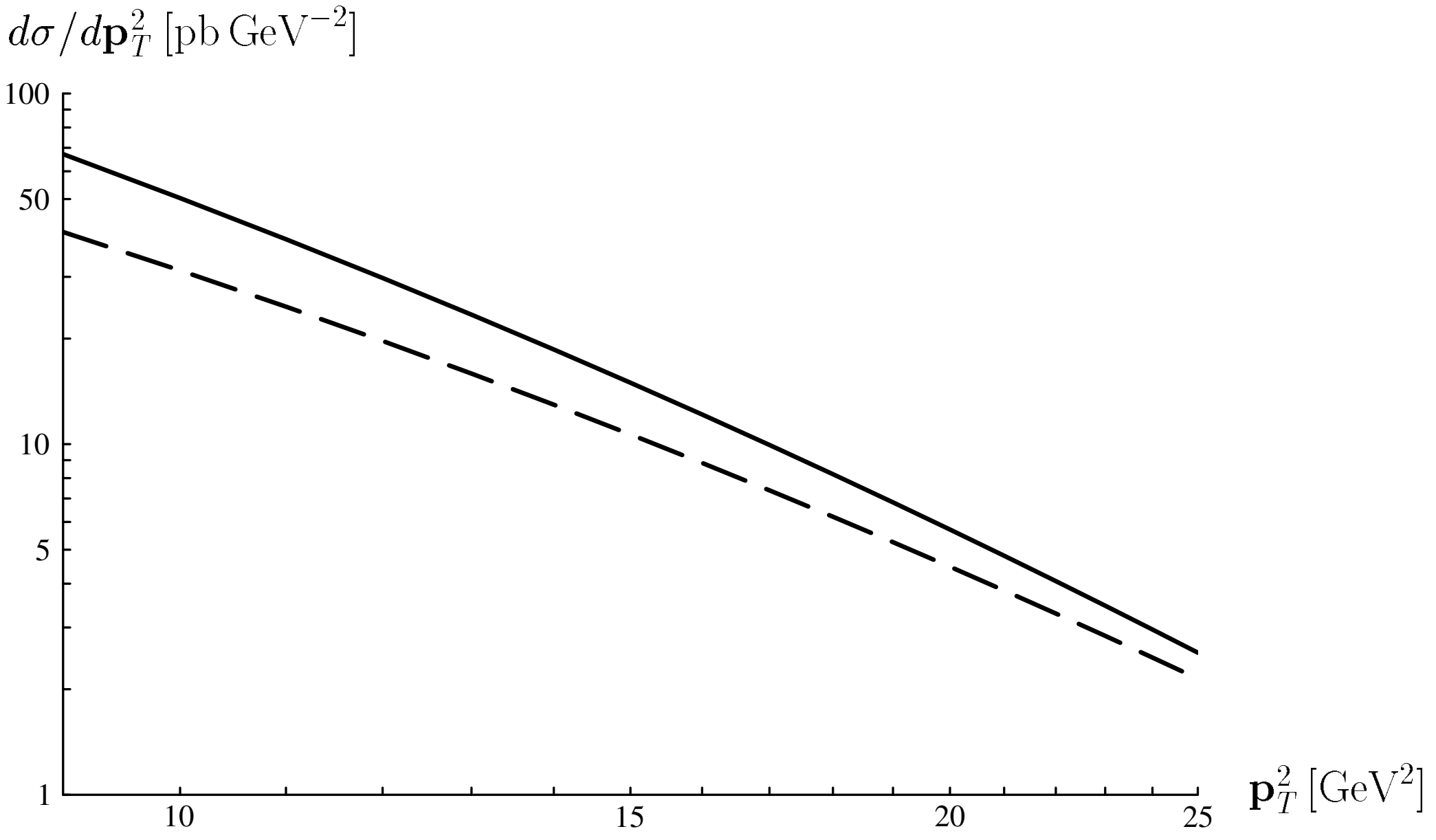

($b$)

Figure 7: ($a$) The $\mathbf{p}_T^2$-spectrum $d\sigma/d\mathbf{p}_T^2$ for $\xi \leq 0.01$, $W = 200\,\mathrm{GeV}$ and $Q^2 = 10\,\mathrm{GeV}^2$. The curves are for the production of $u$, $d$ and $s$ quarks (with current masses) and transverse $\gamma^*$ (full), the same quarks and longitudinal $\gamma^*$ (dash-dotted), and for charm quarks and transverse $\gamma^*$ (dashed). ($b$) The same at $Q^2 = 40\,\mathrm{GeV}^2$, without the curve for the (rather small) longitudinal cross section.

photons and light quarks it increases with $\mathbf{p}_T^2$ and is largest for low $Q^2/\hat{s}$, where the nonperturbative cross section is very small. If one sets $m = 0$ and $\hat{s} = 4\mathbf{p}_T^2$, then for $|\mathbf{p}_T| = 3\,\mathrm{GeV}$ the perturbative tail gives a correction of 10% for $Q^2 = 5\,\mathrm{GeV}^2$ and less than 3% when $Q^2 \geq 15\,\mathrm{GeV}^2$. With $|\mathbf{p}_T| = 6\,\mathrm{GeV}$ one finds an 8% correction at $Q^2 = 40\,\mathrm{GeV}^2$, which grows to 19% at $Q^2 = 15\,\mathrm{GeV}^2$. There however, the longitudinal cross section dominates over the transverse one anyway, so the perturbative correction is not too large at least for values of $Q^2$ where the transverse cross section is important. Also, the absolute size of the perturbative contribution is always small compared with the scale set by the maximum values of the nonperturbative transverse and longitudinal cross sections. As for the production of charm quarks, we have seen that the nonperturbative transverse cross section does not vanish for low $Q^2$. Even at $Q^2 = 0$ the perturbative correction is below 4% for $|\mathbf{p}_T| = 3\,\mathrm{GeV}$, and less than 11% when $|\mathbf{p}_T|$ is twice as large.

In the case of longitudinal photon polarisation the perturbative correction to the cross section is smaller and can be neglected. Where the cross section for longitudinal photons calculated with $\alpha_s^{(0)} D(-\mathbf{l}_T^2)$ vanishes, the contribution from the perturbative tail is unobservably small compared with the transverse cross section.

As remarked in the previous section, the simple substitution of the nonperturbative by the perturbative gluon propagator and the introduction of the running strong coupling for all vertices may not be sufficient to describe the physics at large gluon momenta. If one takes the Lipatov intercept of the pomeron trajectory when the gluons in our one-loop calculation are hard, the contribution from the perturbative tail dominates over the one from small $\mathbf{l}_T^2$ when $\xi$ is some parts in $10^{-3}$ in the case of transverse photons, for longitudinal ones it is larger for some values of the parameters $\mathbf{p}_T^2$, $\hat{s}$, $Q^2$, and smaller for others. We emphasise again that this is not meant as a proper treatment of the perturbative pomeron, but rather as an indication of what the numerical situation is under a certain assumption.

# 7 The pomeron structure function

In the present section we shall use our calculation of diffractive jet production to estimate the quark structure function of the pomeron.

More precisely, one can generalise the usual relation between the cross section for the absorption of transverse photons by the proton and the quark distribution functions $q(x_B)$,

$$\sigma_T(\gamma^* p) = \frac{4\pi^2 \alpha}{W^2 + Q^2} \sum_q e_q^2\, q(x_B) \tag{39}$$

with our choice of $2(W^2 + Q^2)$ for the photon flux factor, to the differ-ential cross section $d\sigma_T/(dt\,d\xi)$ for diffractive photoabsorption, introducing the diffractive distribution functions $dq/(dt\,d\xi)$. Here $x_B = Q^2/(2\nu)$ is the Bjorken variable, and the sum over $q$ is to be taken for quarks and antiquarks. Under the assumption that diffractive scattering is dominated by pomeron exchange, and that the coupling of the pomeron to the proton and the diffractive final state factorises, one can write $dq/(dt\,d\xi)$ as the product of the "flux of pomerons in the proton" and a quark structure function $G_{q/\mathbb{P}}(z)$ of the pomeron. Since the pomeron is an isosinglet, the quark and antiquark distributions are equal. We use the definition of [13], that is

$$x_B \frac{dq(x_B)}{dt\,d\xi} = \frac{9\beta_0^2}{4\pi^2}[F_1(t)]^2\, \xi^{1-2\alpha_{\mathbb{P}}(t)}\, G_{q/\mathbb{P}}(z) \ , \tag{40}$$

where

$$z = x_B/\xi = Q^2/(\hat{s}+Q^2) \tag{41}$$

is the scaling variable for the photon-pomeron subreaction. Together with the diffractive equivalent of (39) this gives

$$\frac{d\sigma_T}{dt\,d\xi} = \frac{\alpha}{\hat{s}+Q^2}\, 9\beta_0^2\,[F_1(t)]^2\, \xi^{1-2\alpha_{\mathbb{P}}(t)} \sum_q e_q^2\, \frac{1}{z} G_{q/\mathbb{P}}(z) \ . \tag{42}$$

In the phenomenological model of the pomeron which is described briefly in the appendix, one obtains a structure function

$$\widetilde{G}_{q/\mathbb{P}}(z) = \frac{3\beta_0^2\mu_0^2}{8\pi^2}\, z(1-z) \approx 0.2\, z(1-z) \tag{43}$$

for each massless quark or antiquark and $z$ not too small, say for $z \geq 0.1$ [13].

Since $p+\gamma \to p+q\bar{q}$ (fig. 1) is the "Born approximation" of diffractive photoabsorption, we may use our present calculation to estimate the pomeron quark structure in the LN model. This requires integration of the differential cross section (22), (25) over $\mathbf{p}_T^2$ from zero to its kinematical upper limit. A hard scale in the photon-pomeron reaction is now provided by $Q^2$, and we shall furthermore require that the squared invariant mass $\hat{s}$ be large enough, say $\hat{s} \geq 20\,\mathrm{GeV}^2$. For fixed $Q^2$ this puts an upper limit on the scaling variable $z$ in (41) for which we assume our calculation to be sensible. However, it can be seen from (21) and (35) that if $\mathbf{p}_T^2$ is small and $\mathbf{l}_T$ is near $\mathbf{p}_T$, the squared momentum of the off-shell quark in diagram ($b$) or ($d$) becomes small for light quarks, so that the perturbative treatment of the quark-photon coupling and the quark propagator in the upper part of the concerned diagram is problematic in a part of the region of integration.

We have calculated $G_{q/\mathbb{P}}(z)$ in $z$-steps of 0.1 with $Q^2$ equal to 5, 10, 20, 50 and $100\,\mathrm{GeV}^2$, the requirement $\hat{s} \geq 20\,\mathrm{GeV}^2$ leading to $z \leq 0.2$, 0.3, 0.5, 0.7 and 0.8, respectively. We further take $z \geq 0.1$, corresponding to $w \geq 10/9$, so that the problems associated with $w \to 1$ which occur for photoproduction do not exist (cf. fig. 4).[13]

When calculating the $\mathbf{p}_T^2$-integrated cross section, we can no longer expand the loop integrand in $\mathbf{l}_T^2/\mathbf{p}_T^2$, after which only a moment of the gluon propagator was needed, and therefore we take the particular form (31) of $D(l^2)$ for different values of $n$. The dependence of our results on $n$ is very weak, of the order of 5%, which gives some hope that the precise shape of $D(l^2)$ is not too important. As for the running coupling $\alpha_s(\mathbf{p}_T^2)$, which we introduced for the vertex involving the off-shell quark, we choose to freeze it when it becomes equal to one, admitting that again this is far from being a rigourous treatment. Using the QCD scale parameter $\Lambda^{(4)}_{\overline{\mathrm{MS}}} = 260\,\mathrm{MeV}$ [19], the freezing point is at 570 MeV and varies by about 100 MeV when the cited errors on $\Lambda^{(4)}_{\overline{\mathrm{MS}}}$ are taken into account.

We find that our results are not very different in size from (43), with a shape that is somewhat shifted towards larger $z$. In the following, we shall give the ratio of $G_{q/\mathbb{P}}(z)$ obtained in the LN model and $\widetilde{G}_{q/\mathbb{P}}(z)$ in (43). Fig. 8 shows this ratio for light quarks. The scaling in $z$ is very good: where we have several $Q^2$ for the same $z$, differences in the structure function are below 1%. Looking at (22) and (25) to (27), we observe that the loop integrals depend on $\hat{s}$ and $Q^2$ only via $w$, and thus directly on $z$, and that the additional $s$-dependence via $(\mathbf{p}_T^2 + m^2)/\hat{s}$ is not very important for small values of $\mathbf{p}_T^2$. In fact, low $\mathbf{p}_T^2$ dominate in the integrated transverse cross section, the contribution from the region $\mathbf{p}_T^2 > \mu_0^2$ being less than 10% for $m = 0$ and less than 20% for $m = 450\,\mathrm{MeV}$. The dominance of low $\mathbf{p}_T^2$, which is already seen in jet production (cf. fig. 7), can be understood from the behaviour of $f_1$, which is increasing in $v = \mathbf{l}_T^2/\mathbf{p}_T^2$. For $f_2$ the situation is less clear-cut, but the corresponding term in the transverse cross section is suppressed by $m^2/(\mathbf{p}_T^2 + m^2)$ and accounts for at most 25% of the result for light quarks. With the integrated cross section being dominated by low $\mathbf{p}_T^2$, its size is influenced by the value we assume for the strong coupling $\alpha_s^{(0)}$ rather than by the scale from which onwards we take the running coupling $\alpha_s(\mathbf{p}_T^2)$. If for example one uses $\alpha_s = 1$ for all values of $\mathbf{p}_T^2$, the result for the structure function grows only by a factor between 1.2 and 1.4 at $m = 0$ and by 1.4 to 1.6 for $m = 450\,\mathrm{MeV}$.

Because we require $\hat{s}$ not to be too small in our calculation, we need to go to rather high $Q^2$ to cover the whole $z$ range (cf. (41)). The scaling we find presumably exists only in a lowest order approximation; it is to be expected that there will be scaling violation as $Q^2$ increases, due to loop corrections to the upper parts of the Feynman diagrams (fig. 3) or, in other words, due

[13] Also, for very small $z$, the contribution from the triple pomeron vertex becomes important [13].

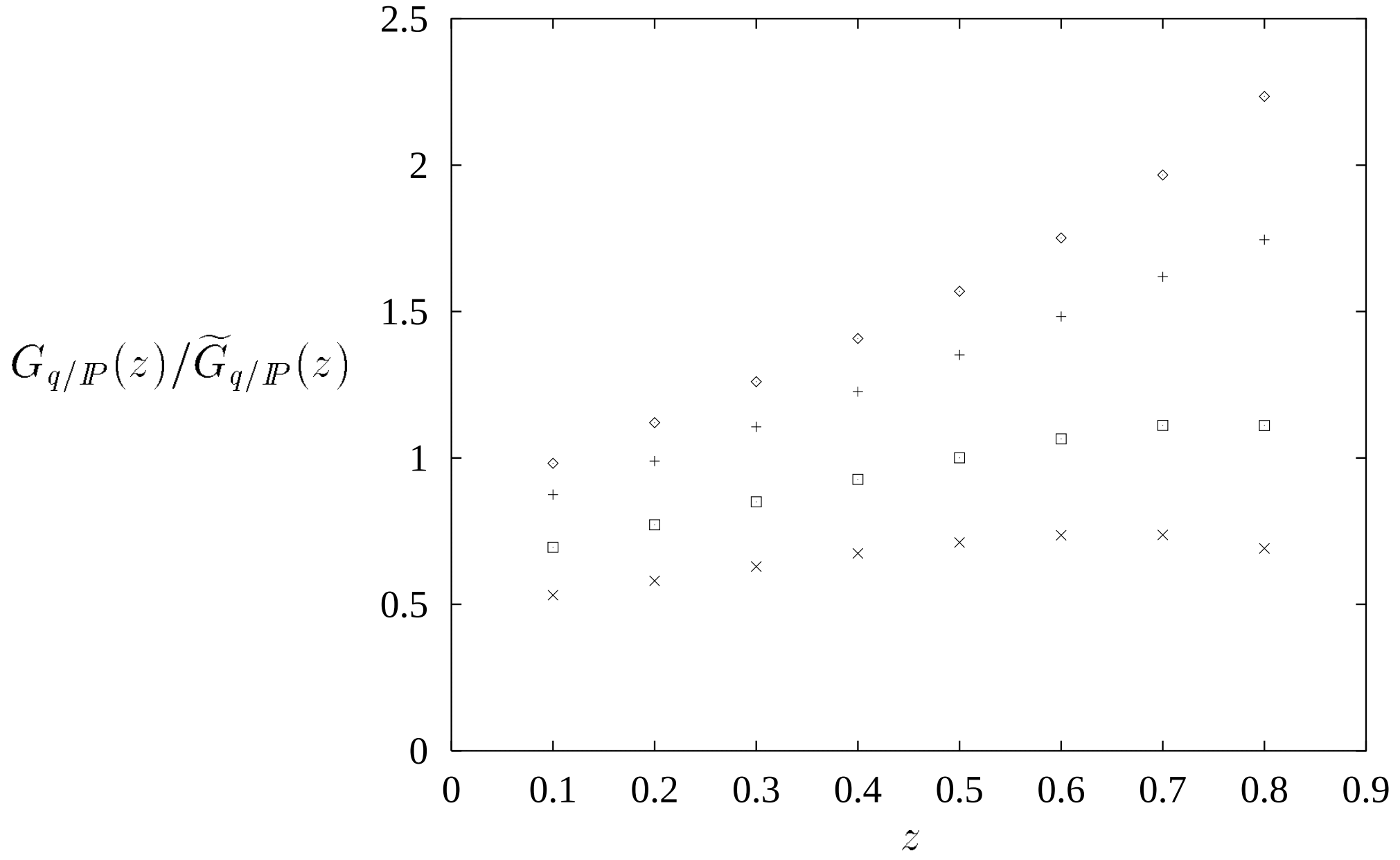

Figure 8: The ratio of the pomeron quark structure function $G_{q/I\!P}(z)$ obtained in the LN model and the result (43) from the phenomenological approach of ref. [13]. The points correspond to $Q^2 = 100\,\text{GeV}^2$ and $m = 0$ ($\diamond$), $m = 150\,\text{MeV}$ (+), $m = 300\,\text{MeV}$ ($\square$), and $m = 450\,\text{MeV}$ ($\times$), respectively. Points for other values of $Q^2$ coincide with the ones shown here to a good precision.

to the perturbative evolution of the structure function.

The dependence on the quark mass is rather pronounced and leads to a flavour asymmetry. The structure function is smaller for $s$ quarks than for $u$ and $d$ quarks, and there is a difference both in shape and in the normalisation according to whether one takes current or constituent masses. Increasing $m$ leads to a suppression at low $\mathbf{p}_T^2$ of the $f_1$-term in (25), due to the factor $\mathbf{p}_T^2/(\mathbf{p}_T^2 + m^2)$, and at the same time increases the parameter $w \propto (\mathbf{p}_T^2 + m^2)/\mathbf{p}_T^2$, so that $f_1$ itself becomes smaller for most values of $v$ (cf. fig. 4).

From $d\sigma_L/(dt\,d\xi)$ one can obtain a longitudinal structure function analogous to (42), which for light quarks turns out to be relatively unimportant. The longitudinal cross section is suppressed at low $\mathbf{p}_T^2$ by a factor $(\mathbf{p}_T^2+m^2)/\hat{s}$, and the longitudinal structure function is less than 6% of the transverse one for $m = 0$. Its importance grows with increasing $m$: at $m = 450\,\text{MeV}$ the ratio of longitudinal and transverse cross section reaches 20% for large $z$, but is still under 6% if one takes out the points at $Q^2 = 50\,\text{GeV}^2$, $z = 0.7$ and at $Q^2 = 100\,\text{GeV}^2$, $z = 0.8$.

For charm quarks, $G_{q/I\!P}(z)$ is about an order of magnitude smaller and suppressed at large $z$ (fig. 9), for the same reasons discussed above in connection with light quark masses. The dominance of low $\mathbf{p}_T^2$ is less pronounced, values of $|\mathbf{p}_T|$ above the charm mass contributing between 12% and 35% to the integrated cross section. As a result, the perturbative region of $\alpha_s$ is now more important (taking a constant $\alpha_s = 1$ gives an answer between 2 and 2.5 times as large). The scaling in $z$ is still quite good: the variation of $G_{q/\mathbb{P}}(z)$ with $Q^2$ is at most 12%, and less than 7% when the values at $Q^2 = 5\,\mathrm{GeV}^2$ are not taken into account. The importance of the longitudinal structure function is no longer negligible at large $z$. It decreases with $Q^2$ at fixed $z$, i. e. it shows no scaling, and increases strongly with $z$. For comparison the longitudinal structure function at two different $Q^2$, normalised to $\widetilde{G}_{q/\mathbb{P}}(z)$, is included in fig. 9.

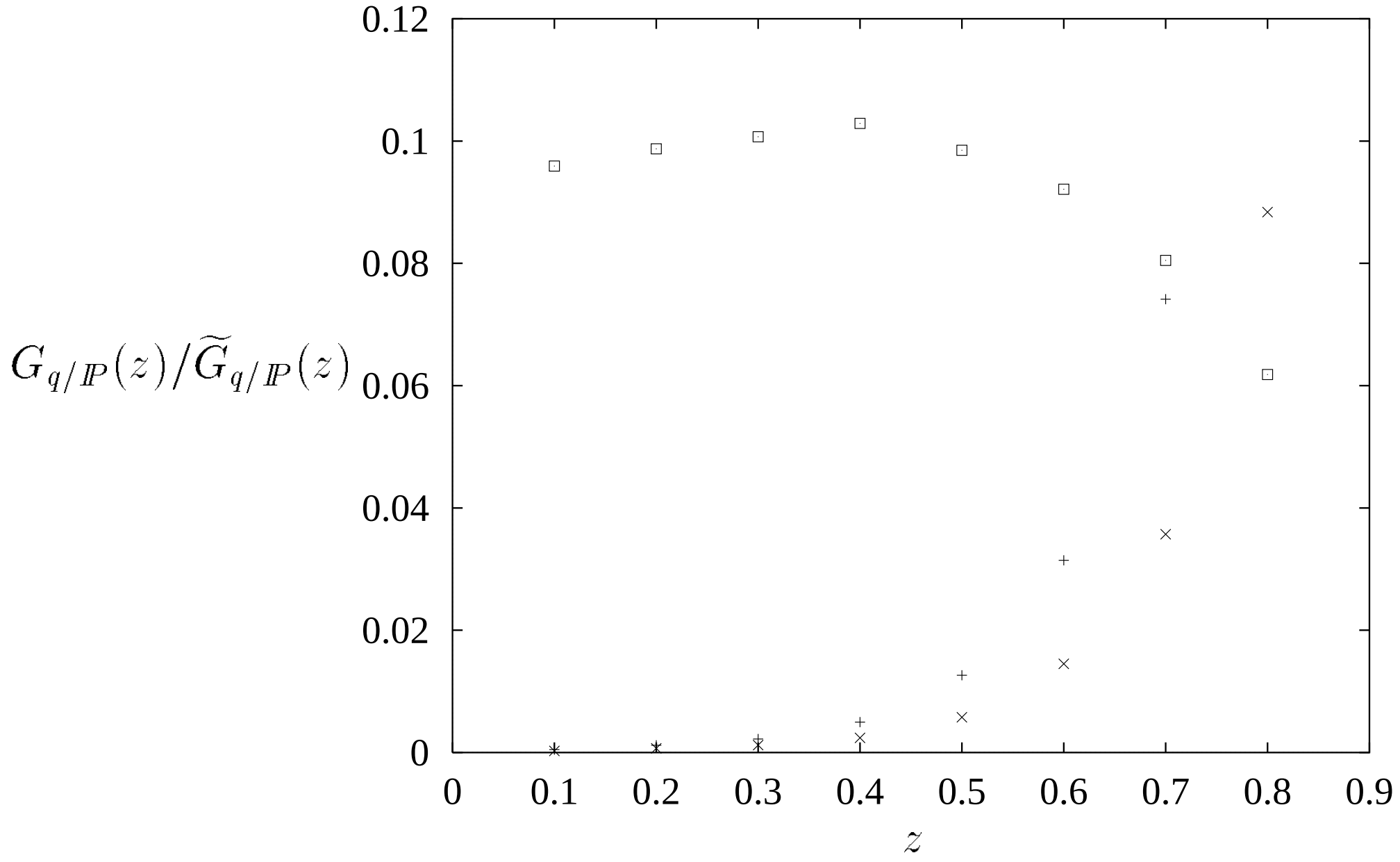

Figure 9: The pomeron structure function for charm quarks divided by (43). Displayed is the transverse structure function at $Q^2 = 100\,\mathrm{GeV}^2$ ($\square$), which is almost the same for other values of $Q^2$. The longitudinal structure function is shown at $Q^2 = 50\,\mathrm{GeV}^2$ (+) and at $Q^2 = 100\,\mathrm{GeV}^2$ ($\times$).

Finally we remark that the contribution of the perturbative tail of the gluon propagator to the structure function can be neglected. The latter is dominated by low $\mathbf{p}_T^2$ and, as we have remarked at the end of section 6, the importance of the perturbative corrections is most important at high $\mathbf{p}_T^2$.

# 8 Conclusions

In the present paper we have calculated the cross section for diffractive production of a quark-antiquark pair with high transverse momentum $p_T$ in photon-proton scattering, modelling the pomeron by two noninteracting gluons with a nonperturbative propagator $D(l^2)$. To a good approximation the result only depends on a moment of $D(l^2)$ which has been estimated from exclusive diffractive $\rho$ production.

We find that in deep inelastic scattering at HERA such events should occur with an observable cross section. The rate for charm quark production is comparable to the one for the light quark flavours, a prediction that appears to us worthwhile testing in experiment. For light quark production there is an important contribution from longitudinally polarised photons at moderate $Q^2$, up to about $20\,\mathrm{GeV}^2$. As expected for soft pomeron exchange, the dependence of the cross section on the photon-proton centre-of-mass energy $W$ is rather weak, the largest effect coming from the reduction of phase space at low $W$, which is due to an upper cutoff we impose on the fractional momentum loss $\xi$ of the scattered proton.

In the case of diffractive *photoproduction* of light quark jets, there is a cancellation in the loop integration unless the gluons are hard, at least if one takes current masses for the produced quarks. To give a nonvanishing contribution, the squared invariant mass of the gluons, which is equal to their squared transverse momentum relative to the pomeron direction, has to be at least of the order of the squared transverse momentum of the jets, and as a consequence the nonperturbative part $D(l^2)$ of the gluon propagator gives an unobservably small cross section. The diagrams we calculate give only the lowest order approximation of pomeron exchange in the process under study. We do not know how corrections from higher loops will modify this result, in particular the replacement of two noninteracting gluons by a ladder structure, which is believed to give an appropriate description of the pomeron. We might also remark that our calculation is done for zero transverse momentum $\Delta_T$ of the diffractively scattered proton, but a nonzero $\mathbf{\Delta}_T^2$ would be small compared with the scale set by the $\mathbf{p}_T^2$ of the final state jets.

For heavy flavour photoproduction the cancellation we have mentioned is not present and we find an appreciable cross section for charm quark jets in our model. If we assume that our result for light quarks is not radically changed by the corrections mentioned above, open charm should be dominating or at least enhanced in diffractive photoproduction. Provided that at least the part of the pomeron which couples to the jets is dominated by large gluon momenta, one might hope to study the *hard* pomeron in the case that experiment finds a significant rate of events with light quark jets, since they cannot be due to purely *soft* pomeron exchange. In any case it would be interesting to study the flavour of the produced jets.

From our calculation of $q\bar{q}$-production we can estimate the total cross section for diffractive photon-proton scattering and extract from it the quark structure function of the pomeron. For this we need a particular ansatz for the gluon propagator $D(l^2)$, but within a class of functions the results are approximately equal. For light quarks we find a structure function $G_{q/\mathbb{P}}(z)$ almost independent of $Q^2$, similar in size to the one obtained in a more phenomenological approach to the pomeron (43) and with a shape somewhat harder than $z(1-z)$. It is amusing to notice that in $p\bar{p}$-collisions, UA8 [20] have measured a pomeron structure function that is harder than $z(1-z)$, although with an uncertainty whether the dominating partons are quarks or gluons. Unlike in jet production, the dependence on the quark mass is rather strong, leading to a smaller structure function for strange quarks than for $u$ and $d$, and to an uncertainty of a factor around 2 according to whether one takes current or constituent masses. For charm quarks the structure function is smaller by about an order of magnitude, and at large $z$ it has a non-negligible contribution from longitudinal photons which is not scaling.

We finally note that some of our results have previously been obtained in the purely perturbative model of Nikolaev and Zakharov [9, 10, 11]. In particular, they have found that the integrated cross section for diffractive $q\bar{q}$-production is suppressed for longitudinal photons and strongly flavour dependent if the photons are transverse, and that at large transverse quark momentum charm is enhanced with respect to light quarks. However, the relative contribution of charm is much smaller in their model than in ours, which was expected in [9]. Furthermore, the part of the quark structure function of the pomeron due to $q\bar{q}$-production was found [10] to behave approximately like $z(1-z)$. As for the production of light quark jets, it was reported [11] that the loop integration is dominated by squared gluon momenta of order $\mathbf{p}_T^2$ if the photon virtuality is very small compared with the invariant $q\bar{q}$-mass, i. e. in particular if the photons are real.

## Acknowledgements

I am grateful to Peter Landshoff for suggesting this work and for many discussions. Thanks are also due to Otto Nachtmann for reading the manuscript, and to Robin Devenish and Peter Bussey for helpful email exchanges. This research is supported in part by the EU Programme "Human Capital and Mobility", Network "Physics at High Energy Colliders", Contract CHRX-CT93-0357 (DG 12 COMA), and in part by Contract ERBCHBI-CT94-1342. It is also supported in part by PPARC.

# A Results in a phenomenological model of the pomeron

In this appendix we give the results of applying the phenomenological model of [13, 14] to the reaction investigated in the present paper.

This approach to the pomeron builds on the observation that it behaves in many ways as an isoscalar photon, and its coupling to light quarks with small invariant mass is taken to be $\beta_0 \gamma_\mu$. If one of the quark legs has large virtuality $k^2$, this coupling has to be multiplied by some form factor $f(k^2)$ to ensure that it vanishes as $k^2 \to -\infty$. The simple ansatz $f(k^2) = \mu_0^2/(\mu_0^2 - k^2)$ has been successful in describing the proton structure function at moderately low $x_B$ and diffractive $\rho$ production in DIS.

In this framework the two diagrams that describe the diffractive process $p + \gamma \to p + q\bar{q}$ are those one obtains from fig. 1 when taking into account the two possible directions of charge flow for the quark line. In the case of zero quark mass $m$, equations (22) to (24) for the cross section then are replaced with

$$\frac{d\sigma}{dt\, d\xi\, d\mathbf{p}_T^2} = \frac{27}{4\pi^2} \alpha e_q^2 \beta_0^4 \, \xi^{1-2\alpha_{I\!P}(t)} \, [F_1(t)]^2 \, \frac{1}{\hat{s} + Q^2} \frac{1}{\sqrt{1 - 4\mathbf{p}_T^2/\hat{s}}} \times$$

$$\left(f(\hat{t}) + f(\hat{u})\right)^2 \times \begin{cases} 1 - 2\,\dfrac{\mathbf{p}_T^2}{\hat{s}} & \text{for transverse photons} \\ \dfrac{\mathbf{p}_T^2}{\hat{s}} \, \dfrac{(\hat{s} - Q^2)^2}{Q^2 \hat{s}} & \text{for longitudinal photons} \\ \dfrac{\mathbf{p}_T^2}{\hat{s}} \, \dfrac{(\hat{s} + Q^2)^2}{Q^2 \hat{s}} & \text{for scalar polarisation} \end{cases} \qquad (44)$$

where by "scalar polarisation" we mean that the polarisation vector of the photon is set proportional to its momentum $q$. We remark in passing that this is very similar to what one would obtain in the LN model considering only diagrams ($a$) and ($c$) in fig. 3 (whose kinematics correspond to that of fig. 1).[14] The above result violates electromagnetic gauge invariance badly: the cross section for scalar photon polarisation is larger than that for longitudinal photons instead of being zero, and both are divergent when $Q^2 \to 0$. For diffractive production of jets with large transverse momentum this model does not give physically reasonable results. Gauge invariance might be restored by introducing an appropriate contact interaction between the photon, the pomeron and the two quarks, as suggested by the field theoretic approach to diffractive scattering of ref. [21].

[14] The cross sections are identical apart from factors $\alpha_s(\mathbf{p}_T^2)/\alpha_s^{(0)}$ in the LN model and $(f(\hat{t}) + f(\hat{u}))^2/4$ in the form factor approach.

If one takes $\mathbf{p}_T^2$ to be small compared with $\hat{s}$, i. e. considers the case where the produced quarks are rather collinear with the photon-pomeron axis, the longitudinal and gauge terms in (44) become small compared with the transverse one and we might believe the results to be more reliable. If we require that $\hat{s} + Q^2 \gg \mu_0^2$ then

$$f(\hat{t}) + f(\hat{u}) \approx \frac{\mu_0^2}{\mu_0^2 + \mathbf{p}_T^2\,(\hat{s} + Q^2)/\hat{s}} \;, \tag{45}$$

and the integrated cross section for the process is dominated by the low $\mathbf{p}_T^2$ region (as it is in the LN model, cf. section 7). Integrating (44) over $\mathbf{p}_T^2 \geq 0$ we find that the results for longitudinal photons and pure gauge are suppressed compared with the transverse one by $\mu_0^2/(\hat{s} + Q^2)$ (apart from logarithms). From the leading term for transverse photons the quark structure function (43) of the pomeron has been extracted in [13].